\documentclass[prd,aps,twocolumn,a4paper,showkeys,nofootinbib]{revtex4-1}

\usepackage{graphicx,psfrag}
\usepackage{mathrsfs}
\usepackage{amsmath,amsfonts,amssymb}
\usepackage{multirow}
\usepackage{comment}

\usepackage{ulem}
\usepackage{hyperref}
\usepackage{enumitem}

\newcommand{\be}{\begin{equation}}
\newcommand{\ee}{\end{equation}}
\newcommand{\bea}{\begin{eqnarray}}
\newcommand{\eea}{\end{eqnarray}}
\newcommand{\bel}{\begin{align}}
\newcommand{\eel}{\end{align}}

\def\l{\ell}
\def\lm{{\ell m}}
\def\p{\partial}

\def\d{{\rm d}}

\def\GMc2{{\rm G M_{\odot} c^{-2}}}

\def\kt2{\kappa^\text{T}_2}

\def\rLR{r_\text{LR}^\text{EOB}}

\usepackage{pifont} % \cmark \xmark

\newcommand{\bajes}{\texttt{bajes}}
\newcommand{\TEOB}{\texttt{TEOBResumS}}
\newcommand{\SEOB}{\texttt{SEOBNRv4}}
\newcommand\gsftides[1]{{\rm GSF{#1}}$^{(+)}${\rm PN}$^{(-)}$}
\newcommand\core[2]{{\tt {#1}:{#2}}}

\usepackage{color}
\definecolor{cyan}{rgb}{0,0.9,0.9}
\definecolor{orange}{rgb}{0.9,0.5,0}
\definecolor{magenta}{rgb}{1,0,1}
\definecolor{purple}{rgb}{0.8,0.4,0.8}
\definecolor{gray}{rgb}{0.8242,0.8242,0.8242}

\begin{document}

\title{Resonant tides in binary neutron star mergers: analytical-numerical relativity study}

\author{Rossella \surname{Gamba}$^{1}$}
\author{Sebastiano \surname{Bernuzzi}$^{1}$}
\affiliation{${}^1$Theoretisch-Physikalisches Institut, Friedrich-Schiller-Universit{\"a}t Jena, 07743, Jena, Germany}

\date{\today}

\begin{abstract}
Resonant excitations of $f$-modes in binary neutron star coalescences
influence the gravitational waves (GWs) emission in both quasicircular
and highly eccentric mergers and can deliver information on the star
interior. Most models of resonant tides are built using approximate, 
perturbative approaches and thus require to be carefully validated 
against numerical relativity (NR) simulations in the high-frequency regime.
We perform detailed comparisons between a set of high-resolution NR
simulations and the state of the art effective one body (EOB) model
{\tt TEOBResumS} with various tidal potentials and including a model
for resonant tides. 
For circular mergers, we find that $f$-mode resonances can improve the
agreement between EOB and NR, but there is no clear evidence that the tidal
enhancement after contact is due to a resonant mechanism. 
Tidal models with $f$-mode resonances do not
consistently reproduce, at the same time, the NR waveforms and the
energetics within the errors, and their performances is comparable to
resummed tidal models without resonances. 
For highly eccentric mergers, we show for the first time that our EOB
model reproduces the bursty NR waveform to a high degree of
accuracy. However, the considered resonant model does not capture the
$f$-mode oscillations excited during the encounters and present in
the NR waveform.
Finally, we analyze GW170817 with both adiabatic and dynamical tides
models and find that the data shows no evidence in favor of models including dynamical tides. 
This is in agreement with the fact that resonant tides are measured
at very high frequencies, which are not available for GW170817 but might be 
tested with next generation detectors.
\end{abstract}


\maketitle

\section{Introduction} 

Tidal resonances in coalescing compact binaries have been studied
for a long time in connection to gravitational-wave (GW) 
observations~\cite{Lai:1993di,Reisenegger:1994,Kokkotas:1995xe,Ho:1998hq}
(See also
\cite{Mashhoon:1973,Mashhoon:1975,Mashhoon:1977,Turner:1977b} for
earlier work on tidally generated radiation.)
During the coalescence process, the proper oscillation modes of a
neutron star (NS) can be resonantly excited by the orbital frequency.
For a quasicircular orbit, the energy transfer between the orbit and
the mode can change the rate of inspiral and alter the phase of the 
chirping GWs~\cite{Ho:1998hq}.
In general, the impact of the tidal resonance on the GWs depends on
the duration of the resonance, and it is stronger the slower the
orbital decay is.
Initial studies focused on the excitation $g$-modes at frequencies
${\lesssim}100\,$Hz, although the effect was found negligible due to
the weak coupling between the mode and the tidal
potential~\cite{Lai:1993di}.
In contrast, $f$-modes have stronger tidal coupling but also higher
frequencies of order ${\sim}(GM_A/R_A^3)^{1/2}$ ($M_A$ and $R_A$ are
the mass and radius of star $A$, respectively), that correspond to a 
few kilo-Hertz for typical NS masses and radii. These frequencies are too
large for the resonance to occur during the inspiral~\cite{Ho:1998hq}; 
their value actually approaching (or being larger than) the merger frequency~\cite{Bernuzzi:2014kca}.

Numerical-relativity (NR) simulations of quasicircular neutron star
mergers conducted so far do not show decisive evidence for the presence of
$f$-mode resonances. One the one hand, some GW models including
$f$-mode resonances have been shown to reproduce the NR waveform
phasing near merger \cite{Hinderer:2016eia,Dietrich:2017feu,Steinhoff:2021dsn}.
On the other hand, the same data can be reproduced at the same
accuracy without assuming the presence of a $f$-mode resonance nor
additional parameters \cite{Bernuzzi:2015rla,Dietrich:2017feu,Nagar:2018zoe,Akcay:2018yyh}. 
Moreover, it is well known that the two
NSs come in contact well before the resonance condition is met
\cite{Thierfelder:2011yi,Bernuzzi:2012ci} (see also discussion below in Sec.~\ref{sec:model}). 
Interestingly, $f$-mode excitation is instead observed in NR simulations of
highly eccentric compact binaries composed of black-hole--NS
\cite{East:2011xa} and two NSs
\cite{Gold:2011df,Chaurasia:2018zhg}. In these mergers, each close
passage triggers the NS's oscillation on proper modes;
the GW between two successive bursts (corresponding to the passages)
clearly shows $f$-mode oscillations (see Fig.~\ref{fig:wfs_ecc} below). 
Note however that the excitation does not meet the resonant
condition~\cite{Gold:2011df}: the close periastron passage exerts a 
tidal perturbation which excites the axisymmetric ($m=0$) $\ell=2$  
mode \cite{Turner:1977b}.

Recent studies after GW170817~\cite{TheLIGOScientific:2017qsa,
  Abbott:2018wiz,Abbott:2018exr} re-considered waveform models 
with $f$-mode resonances and demonstrated the possibility of GW
asteroseismology with binary neutron star inspiral signals
\cite{Pratten:2019sed,Ma:2020rak,Pratten:2021pro}. In 
particular, the prospect study in Ref.~\cite{Pratten:2021pro}
demonstrates that neglecting dynamical tidal effects associated with
the fundamental mode could lead to systematic biases in the inference of
the tidal polarazibility parameters and thus the NS equation
of state. Since GW analyses are performed with matched filtering, these studies
{\it postulate} the validity of resonant models to merger or contact and a
sufficient accuracy of the GW template. While it is, in principle, 
possible to observationally verify the necessity of a
$f$-mode resonance model in a particular observation (e.g. via
hypothesis ranking), the quality of 
current GW data {\it and} templates at high-frequencies is still
insufficient \cite{Gamba:2020wgg}. 
The potential relevance of resonant tides for GW astronomy and the
above considerations motivates further detailed comparisons between the current 
analytical results and numerical relativity simulations.   

In this work, we consider state-of-art models for the compact binary
dynamics with tidal resonances in the effective-one-body (EOB)
framework and critically assess their validity against numerical relativity data. 
In Sec.~\ref{sec:model} we briefly summarize the effective Love number model proposed in
Refs.~\cite{Hinderer:2016eia,Steinhoff:2016rfi} (see also App.~\ref{app:keff}) that can be
efficiently coupled to 
any EOB implementation to generate precise inspiral-merger
waveforms. This model prescribes a dynamical Love number (or tidal 
coupling constant) as function of the quasi-circular orbital frequency
that, while approaching merger, enhances the effect of tidal
interaction. Qualitatively, this effect is known also from
studies of tidally interacting compact binaries with affine models
\cite{Carter:1985a,Luminet:1985a,Wiggins:1999te,Ferrari:2008nr,Ferrari:2011as}.
On physical ground, tidal interactions stronger than those expected by
adiabatic and post-Newtonian models are expected towards merger~\cite{Damour:2012yf}.
For example, 
early EOB/NR comparisons for quasi-circular mergers found that the
description of tidal effects after contact and towards merger requires 
to enhance the attractive character of the 
EOB tidal potential in post-Newtonian form \cite{Bernuzzi:2012ci,Bernuzzi:2013rza,Bernuzzi:2015rla}.
In these studies, it was also pointed out that a key diagnostic to robustly
assess tidal effects in NR data is the use of gauge-invariant energetics \cite{Damour:2011fu}. 

In Sec.~\ref{sec:comp}, we compare different EOB tidal models to
selected, high-resolution NR simulations considering both energetics
and GW phasing. In particular, in Subsec.~\ref{sbsec:circ} we consider 
quasi-circular mergers and show that the $f$-mode resonance does
not give a cosistently accurate description of both energetics and
the waveform.
Similarly, in Subsec.~\ref{sbsec:hyp} we consider a highly eccentric merger and 
show that a $f$-mode resonance model does not qualitatively capture the
``free-oscillation'' feature observed in the frequency-evolution of the 
NR waveform. Notably however -- modulo this effect -- the EOB waveform and (orbital) frequency
closely follow the NR quantities up to $\sim$ one orbit before merger, 
attesting to the goodness of the dynamics description provided by the model even for these extreme systems.

In Sec.~\ref{sec:GW170817} we perform Bayesian analyses and model
selection on GW170817 data using the various EOB models introduced in Sec.~\ref{sec:model}.
We find that $f$-mode augmented models are not favored with respect
to models which only implement adiabatic tidal effects. 
The $f$-mode resonant frequencis cannot be measured in GW170817, as
also observed in \cite{Pratten:2019sed}.
This is expected, since $f$-mode inference is mostly informative at
frequencies larger than ${\sim}1$kHz (for comparable and canonical NS 
masses), and GW170817 may not contain enough high-frequency 
information to allow for such a measurement.

Finally, in Sec.~\ref{sec:conc} we 
conclude that, while the $f$-mode model can be effective in improving the 
agreement between NR and EOB after contact and to merger, it
is not clear whether this corresponds to the actual resonant effect or
if it is rather an effective description for
the hydrodynamics-dominated regime of the merger.  
Hence, caution should be applied whenever trying to extract actual
physical information (i.e., the $f$-mode resonant frequencies) 
from a matched filtered analysis using templates
that include $f$-mode resonances.

\paragraph*{Notation.} We use geometrical units, $c=G=1$.
We indicate the total binary mass as $M=m_1+m_2$, the reduced mass as
$\mu=m_1m_2/M$, the mass fraction of star $A$ as $X_A=m_A/M$, the mass
ratio is $q=m_1/m_2=X_1/X_2\geq1$ and the
symmetric mass ratio as $\nu=\mu/M$. The EOB variables are mass
rescaled, 
\begin{align}
  r = R/(GM) \ , \ \
%  u = 1/r \ , \ \
  t = T/(GM)\ , \ \
  \hat\Omega = \frac{d\varphi}{dt} \ .
\end{align}
In these variables Kepler's law is $\hat\Omega^2 = u^3$ with $u=1/r$.

Given the dimensionless Love number for star $A$, $k^{\rm (A)}_\ell$, the
tidal polarizability parameters are defined as
\be
\Lambda^{\rm (A)}_\ell = \frac{2}{(2\ell-1)!!} C^{2\ell+1}_A k^{\rm (A)}_\ell  \ ,
\ee
where $C_A=GM_A/(c^2R_A)$. The tidal coupling constants of star $A=1$ are given by
\be
\kappa^{\rm (1)}_\l = (2\l-1)!!\, \Lambda^{\rm (1)}_\l \frac{X_1^{2\l}}{X_2} \, ,
\ee
and $\kappa_2^{\rm T} = \kappa_2^{(1)}+\kappa_2^{(2)}$.
The $f$-mode frequency with index $\ell$ of
star $A$ is indicated as $\omega^{\rm (\ell)\,A}_f$, and we use 
$\bar\omega^{\rm (\ell)}_{f\,{\rm A}}=Gm_{\rm A}\omega^{\rm
  (\ell)}_{f\,{\rm A}}$ (the labels $A$ and $(\ell)$
are sometimes dropped).

\section{Effective Love number model}
\label{sec:model}

\begin{figure}[t]
  \centering 
  \includegraphics[width=0.49\textwidth]{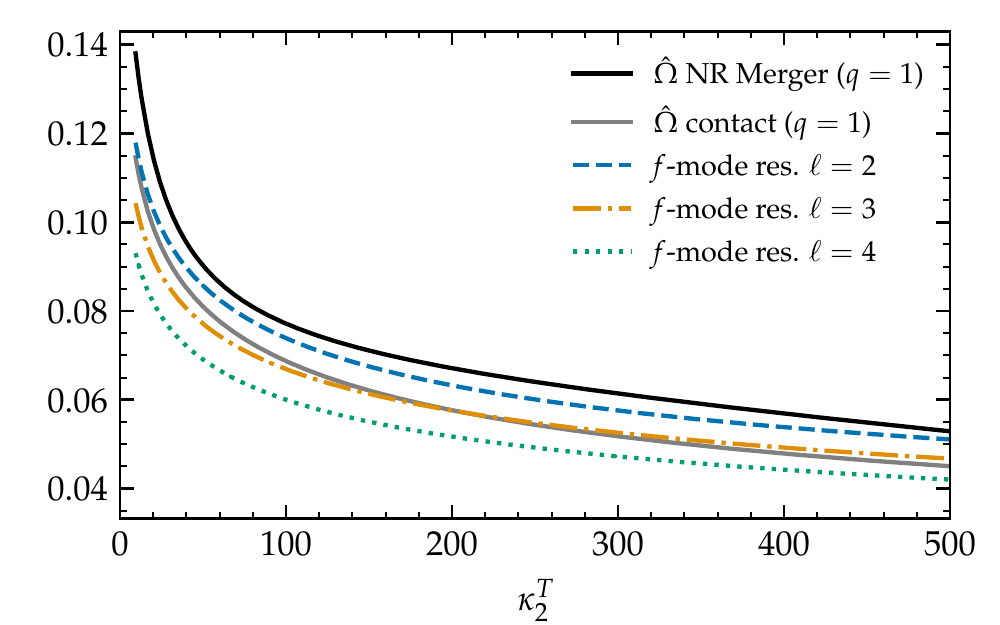}
  \caption{Mass-rescaled orbital merger frequency, contact frequency
    and resonant conditions for $\ell=2,3,4$ modes for equal-mass
    binaries with different tidal coupling constant $\kappa_2^{\rm T}$.
    The merger frequency is computed from the NR quasiuniversal fits
    of Ref.~\cite{Bernuzzi:2014kca}. The contact frequency is
    estimated as in \cite{Damour:2009wj} using the quasiuniversal
    relations $C_A(\Lambda_2^{\rm (A)})$ of \cite{Godzieba:2020bbz}.} 
  \label{fig:Mrg_f_freqs}
\end{figure}

The model of Refs.~\cite{Hinderer:2016eia,Steinhoff:2016rfi} describes the
resonant excitation of the NS $f$-mode by a {\it circular} orbit based
on an {\it effective} quadrupolar Love number. The latter is defined
by an approximate, Newtonian solution of
\be\label{eq:keff_def}
k^{\rm eff}_2 = \frac{E_{ij}Q^{ij}}{E^2}\,, 
\ee
where $E_{ij}=\p_i\p_j\phi$ is the external quadrupolar field derived from
the Newtonian potential $\phi$ and $Q_{ij}$ is the NS's quadrupole.
The resonance of a NS's modes is triggered by the condition
\be\label{eq:resonance}
m\hat\Omega\,X_{\rm A} = \bar\omega^{\rm (\ell)}_{f\,{\rm A}}\,, 
\ee
and its net effect is an enhancement of the Love number $k^{\rm (A)}_\ell$.
This results in a simple prescription to obtain ``dynamical tides''
based on the formal subsitution of the Love numbers (or
equivalently the tidal coupling constants) with their effective values: 
\be\label{eq:keff}
k^{\rm (A)}_\ell\mapsto k_\ell^{\rm eff\,(A)}:=\alpha_{\ell
  m}(\nu,\hat\Omega,\bar{\omega}_{f\,{\rm A}}^{(\ell)},X_{\rm A})k_\ell^{\rm (A)} \,
\ee
where the dressing factor $\alpha_{\ell m}$ in Eq.~\eqref{eq:keff} is
a multipolar correction valid for $\ell=m$. All the expressions are
given in Appendix~\ref{app:keff}.

In this work, the model with $\ell=2,3,4$ resonances is incorporated
in \TEOB{} (\verb#v3# ``GIOTTO'')~\cite{Damour:2014sva,Nagar:2015xqa,Nagar:2018zoe,Nagar:2019wds,Akcay:2020qrj,Nagar:2020pcj,Gamba:2020ljo,Riemenschneider:2021ppj}. Tidal
interactions are described by additive contribution
$A_{\rm T}$ to the EOB metric potential
\cite{Damour:2009wj}. Different choices for $A_{\rm T}$ are considered:
(i) a post-Newtonian (PN) baseline expression including NNLO
gravitoelectric corrections \cite{Bernuzzi:2012ci} 
(as also employed in Refs.~\cite{Hinderer:2016eia,Steinhoff:2016rfi})
and LO gravitomagnetic terms \cite{Akcay:2018yyh},
(ii) a resummed expression of high-order gravitoelectric $\ell=2$ PN terms obtained from
gravitational self-force computations
\cite{Bini:2014zxa,Bernuzzi:2015rla}, hereafter referred as
\gsftides{2} (See Tab.I of \cite{Akcay:2018yyh});
(iii) a resummed expression of high-order gravitoelectric $\ell=2,3$ PN terms obtained from
gravitational self-force computations (\gsftides{23}).
\TEOB{}'s GIOTTO release also includes the LO gravitoelectric PN terms up to
$\ell=8$ \cite{Godzieba:2020bbz,Godzieba:2021vnz}. Tidal terms in the
other EOB potentials and in the waveform are described in detail in Ref.~\cite{Akcay:2018yyh}.

For typical binaries the resonant condition in Eq.~\eqref{eq:resonance} is
met before the moment of merger (defined as the peak of the $\ell=m=2$
mode of the strain). This is shown in Fig.~\ref{fig:Mrg_f_freqs} for
equal-mass mergers, where the merger frequency (solid black line) is
computed in terms of the tidal coupling constant $\kappa_2^{\rm T}$
using the quasiuniveral relations of Ref.~\cite{Bernuzzi:2014kca}.
The contact frequency (gray solid line) is estimated as in Eq. (78) of
\cite{Damour:2009wj}; this simple expression is known to 
overestimate the values extracted from the
simulations\footnote{A better representation
would be obtained accounting also for the shape love number of the stars, $h$ \cite{Damour:2009wj,Bernuzzi:2012ci}} -- e.g. $\hat\Omega\sim 0.04$ for equal mass NSs with
$\kappa_2^{\rm T}\sim 180$, 
effectively corresponding to the
last 2-3 GW cycles to the moment of merger \cite{Bernuzzi:2012ci} --
but provides a sufficient estimate for this work.

Colored (non-solid) lines indicate that the resonant excitation for
the $\ell=2,3,4$ $f$-mode happens progressively earlier in the merger
process. While the $\ell=2$ $f$-mode is excited shortly before merger
(approximately corresponding to the last GW cyles) and after
contact, the octupolar and hexapolar $\ell=3,4$ mode resonances are 
reached before the NSs' contact.
This has two important implications. First, the predicted resonance
phenomenon can be directly tested with numerical relativity
simulations and should, if significant, be visibile in the
gauge-invariant energetics of the dynamics from the simulations.   
Second, the dominant $\ell=2$ resonance happens in a regime in which
the model itself is not valid since the NSs are not anymore isolated
nor ``orbiting''; the matter dynamics being governed by hydrodynamical
processes.

\begin{figure}[t]
  \centering 
  \includegraphics[width=0.49\textwidth]{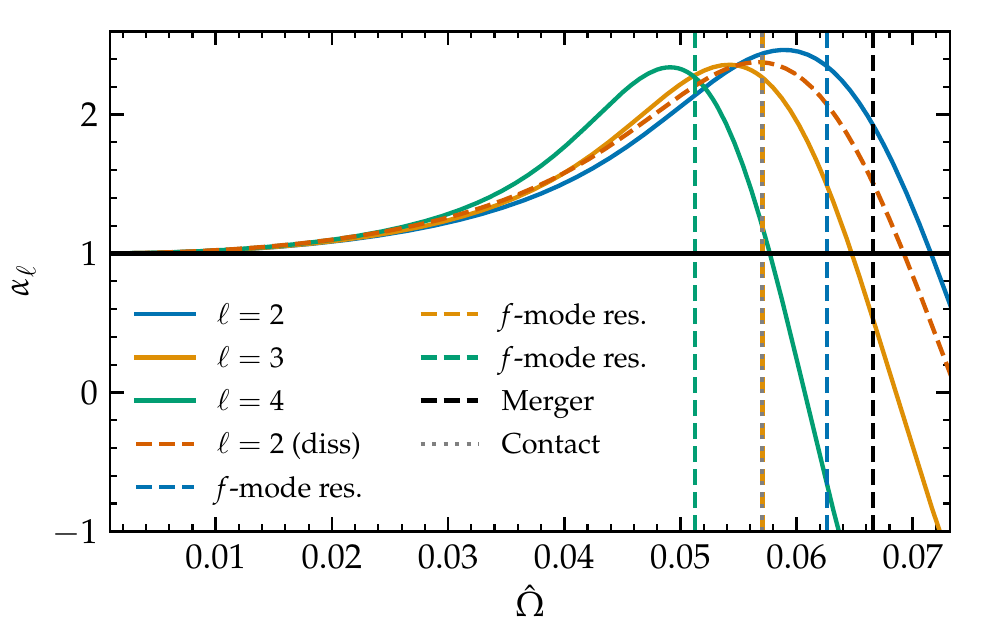}
  \caption{Dressing factors prescribed by the effective Love number model
    as a function of the orbital frequency and for a fiducial
    binary. Vertical lines mark the resonances and the merger
    frequency. The explicit expressions for $\alpha_\lm$ are given in
    Appendix~\ref{app:keff}. The dashed cyan line is the waveform's
    amplitude correction of Eq.~\eqref{eq:hathalpha22}. The dotted gray line,
    which happens to be superimposed to the green $\ell=3$ resonance, corresponds
    to the contact frequency of the stars.} 
  \label{fig:keff}
\end{figure}

\begin{figure*}[t]
  \centering 
  \includegraphics[width=0.49\textwidth]{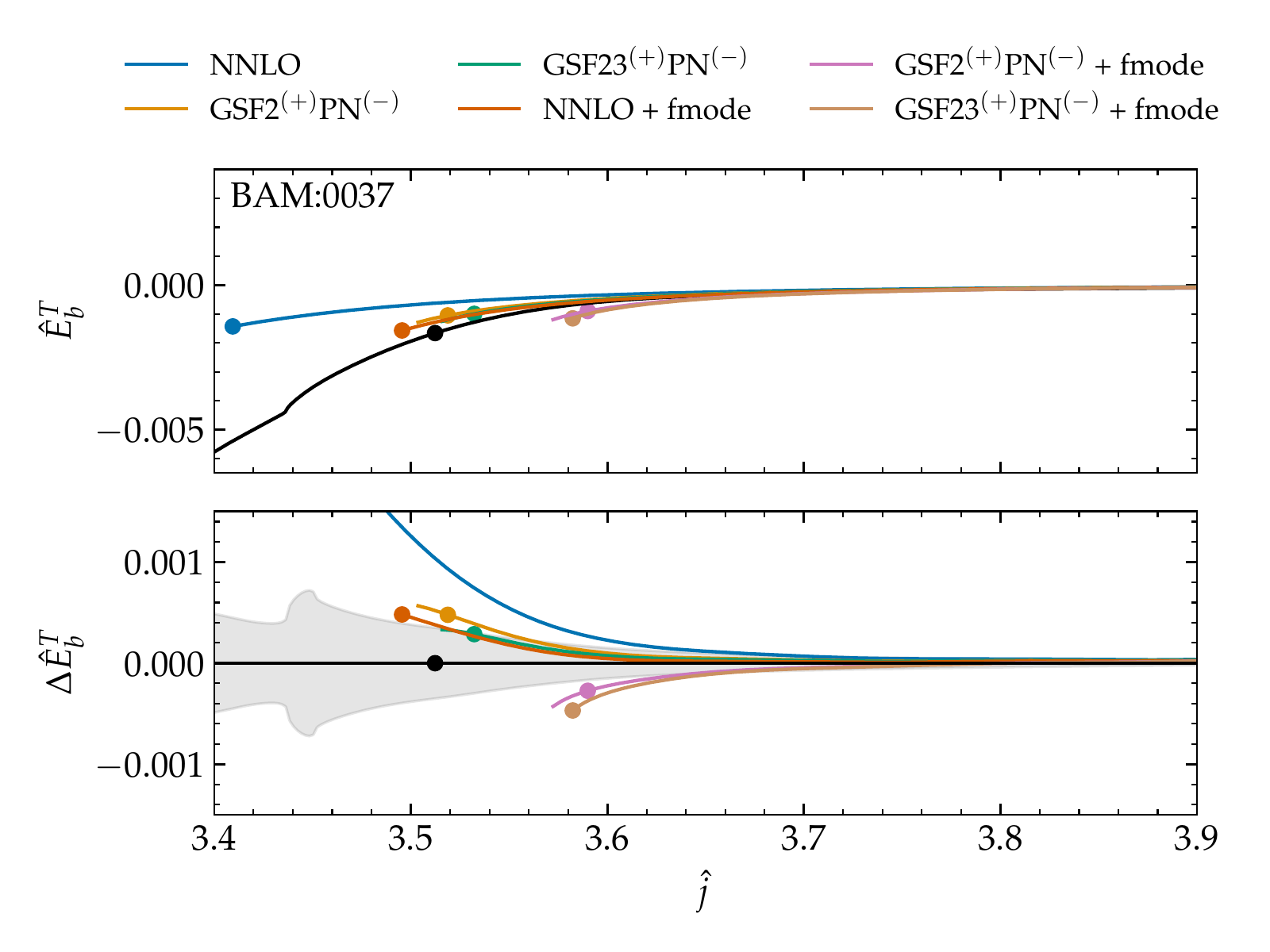}
    \includegraphics[width=0.49\textwidth]{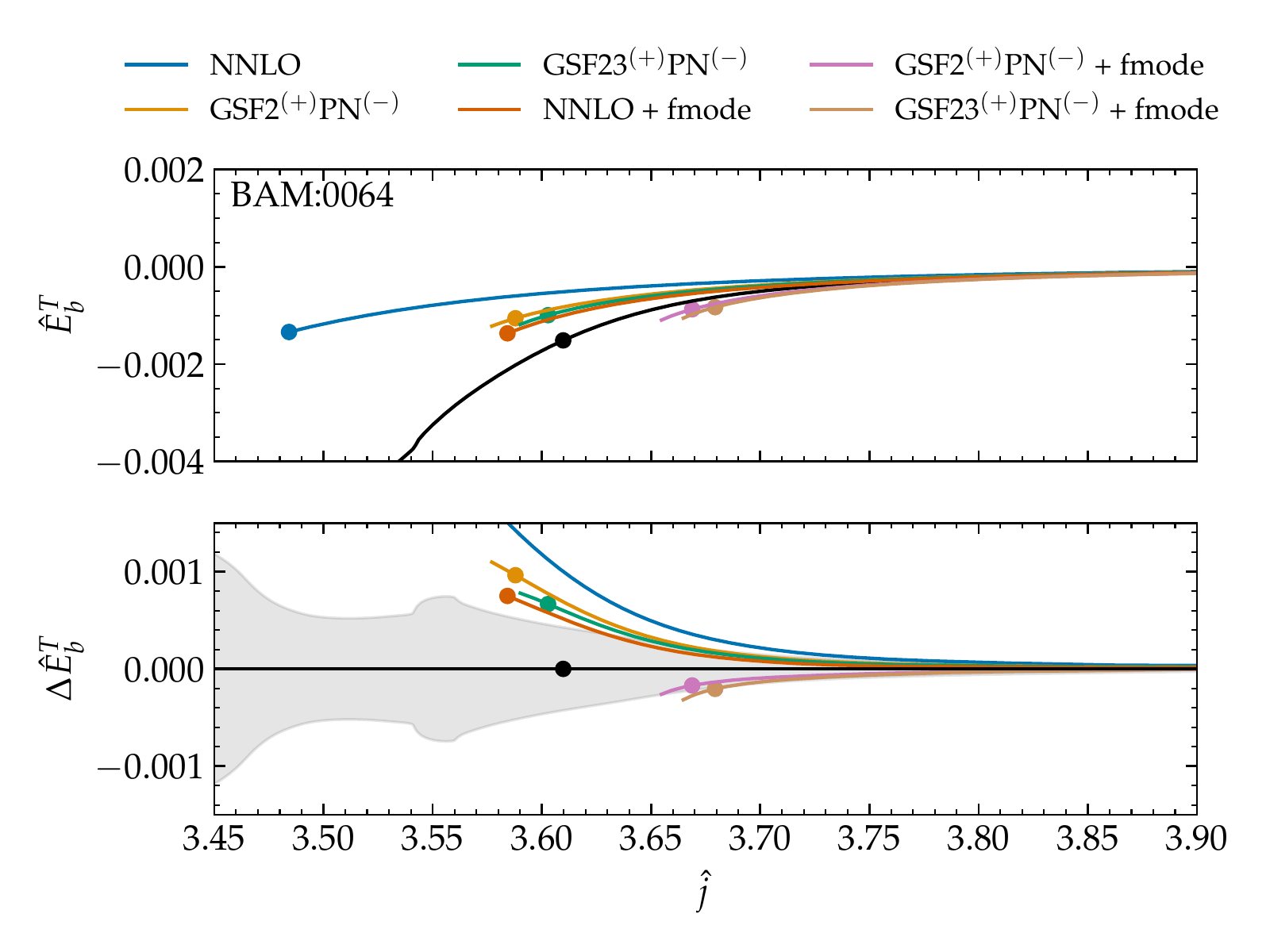}
  \includegraphics[width=0.49\textwidth]{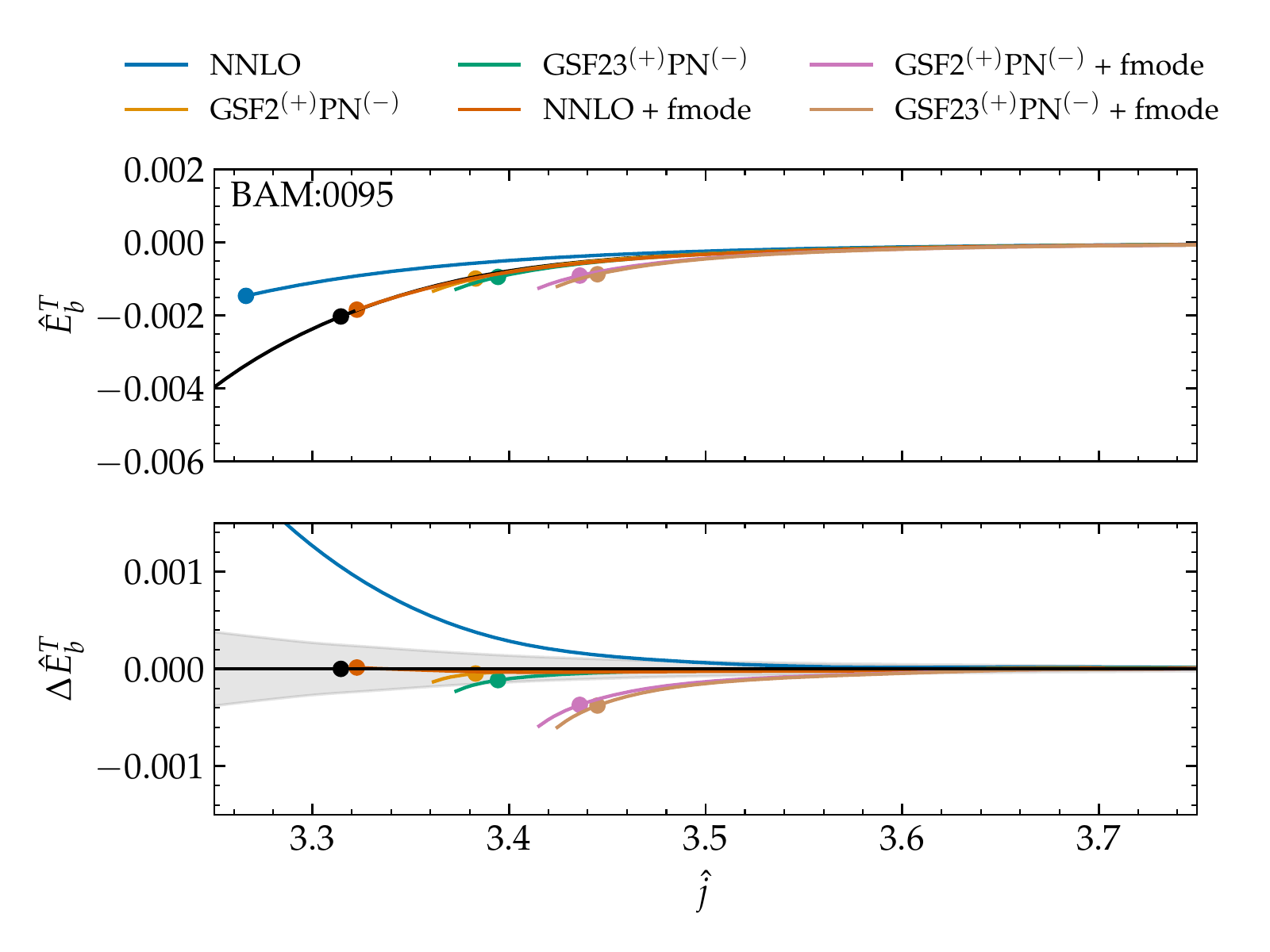}
  \includegraphics[width=0.49\textwidth]{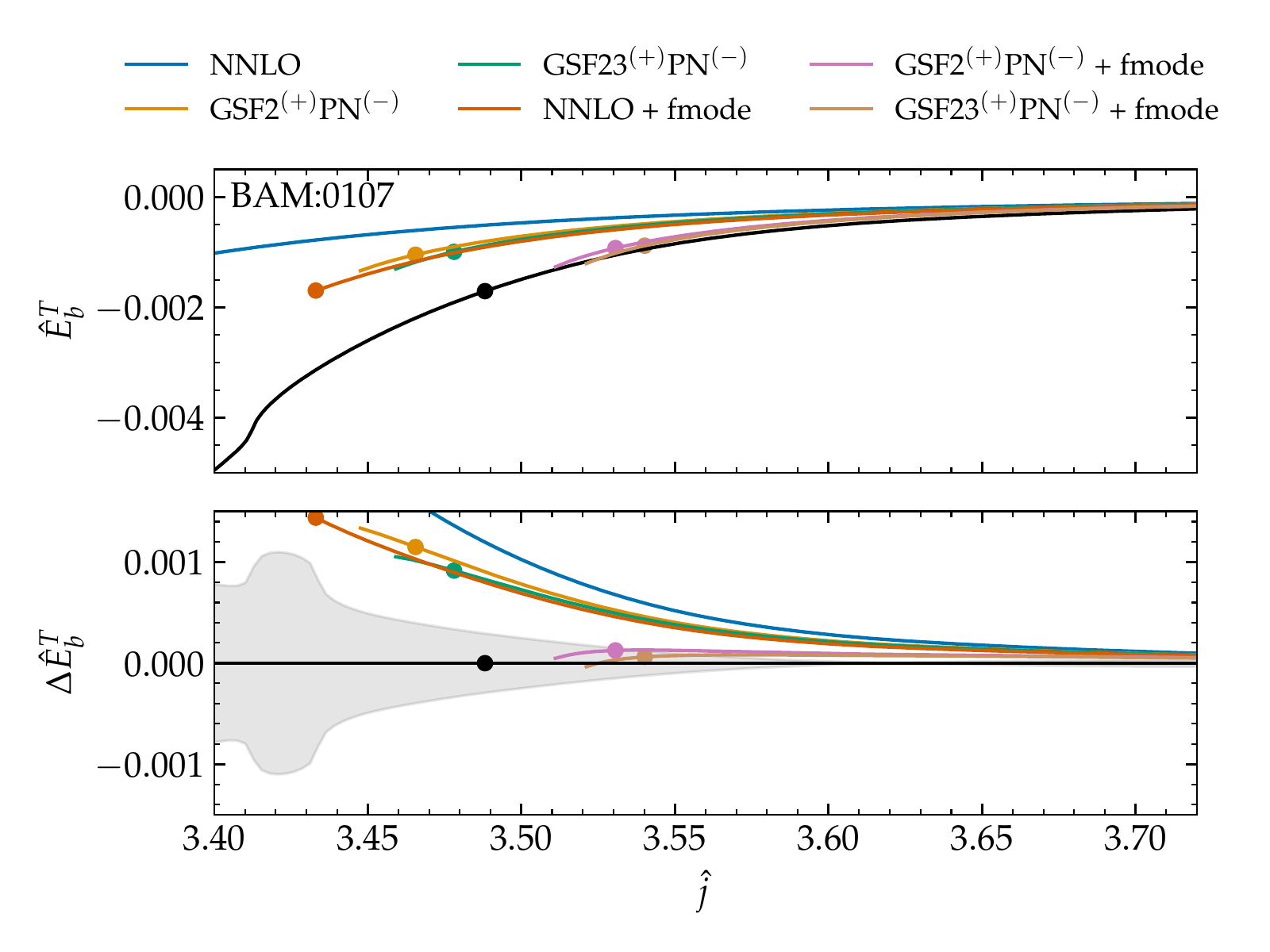}
  \caption{$\hat{E}_b^T$ and $\Delta\hat{E}^T_b$
  		 as functions of the angular momentum $\hat{j}$ of the system for
  			the equal mass BAM simulations considered in this paper (black) and
  			the respective \TEOB~ simulations. The latter are computed using
  			different baseline tidal models (NNLO, GSF2${}^{(+)}$PN${}^{(-)}$, GSF23${}^{(+)}$PN${}^{(-)}$)
  			and f-mode contributions.
			The EOB and NR mergers are denoted via dots, while shaded gray bands indicate the NR
  			error.}
  \label{fig:EbjT_qcs}
\end{figure*}

\begin{figure*}[t]
  \centering 
  \includegraphics[width=0.45\textwidth]{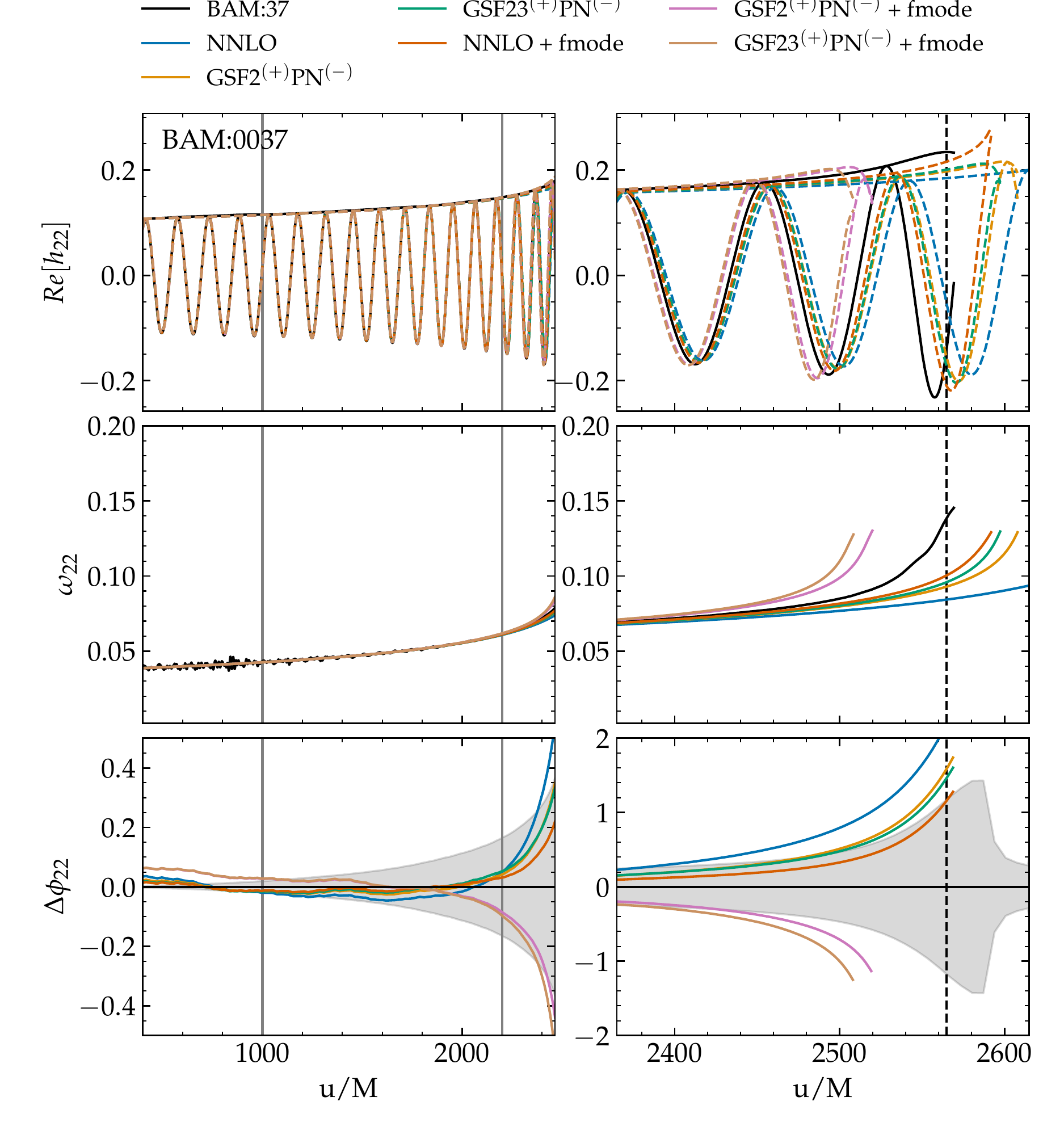}
  \includegraphics[width=0.45\textwidth]{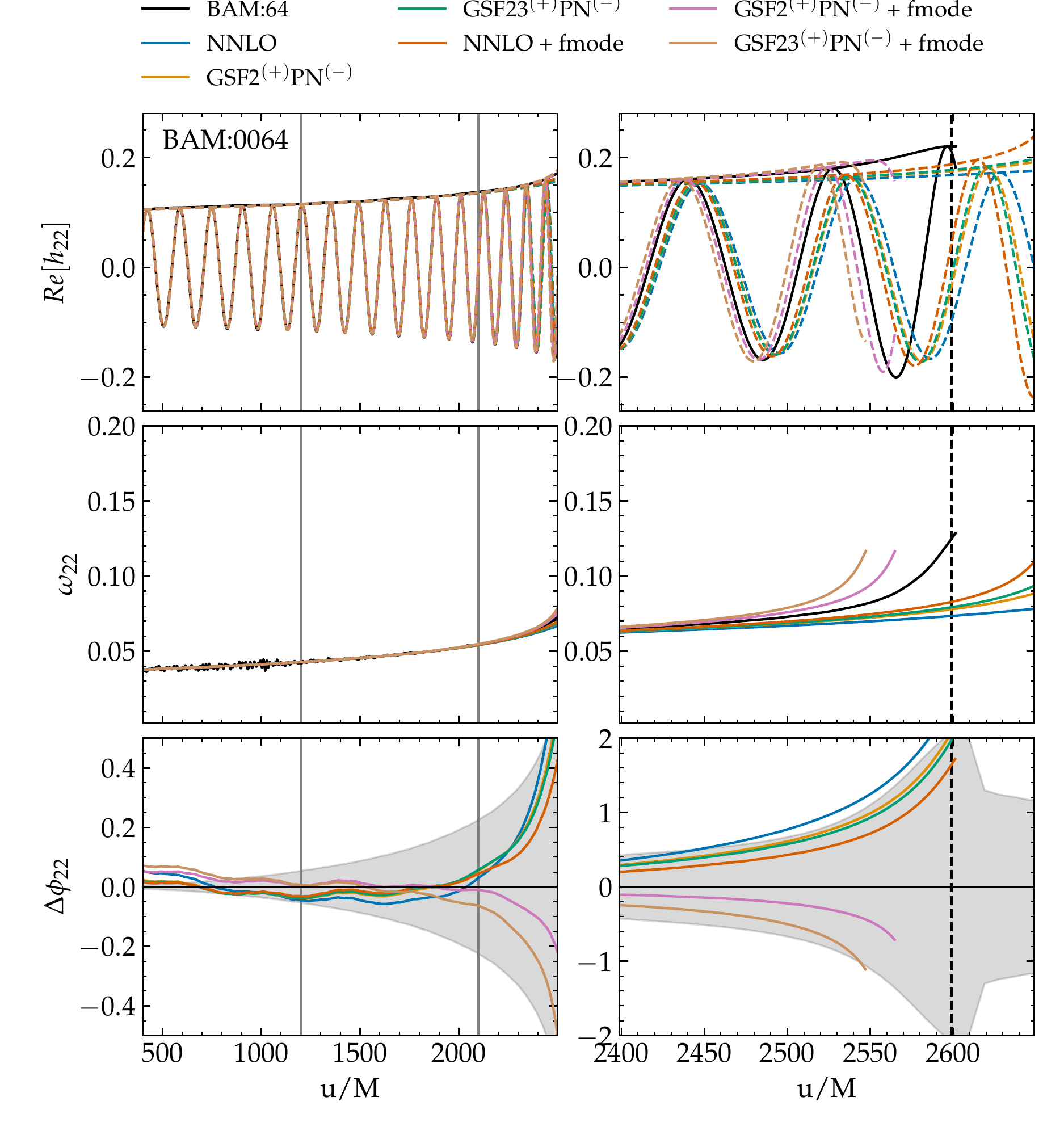}
  \includegraphics[width=0.45\textwidth]{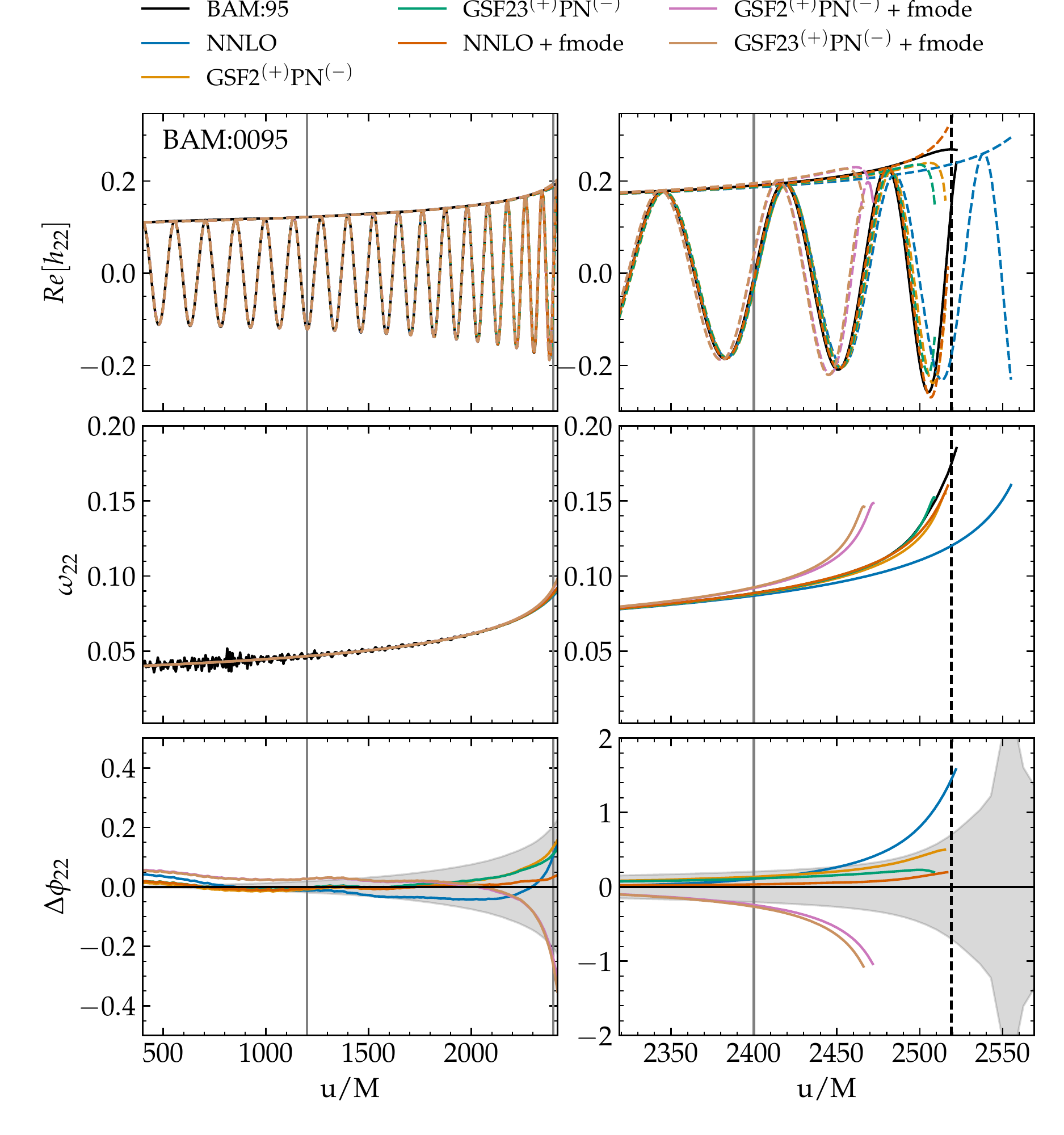}
  \includegraphics[width=0.45\textwidth]{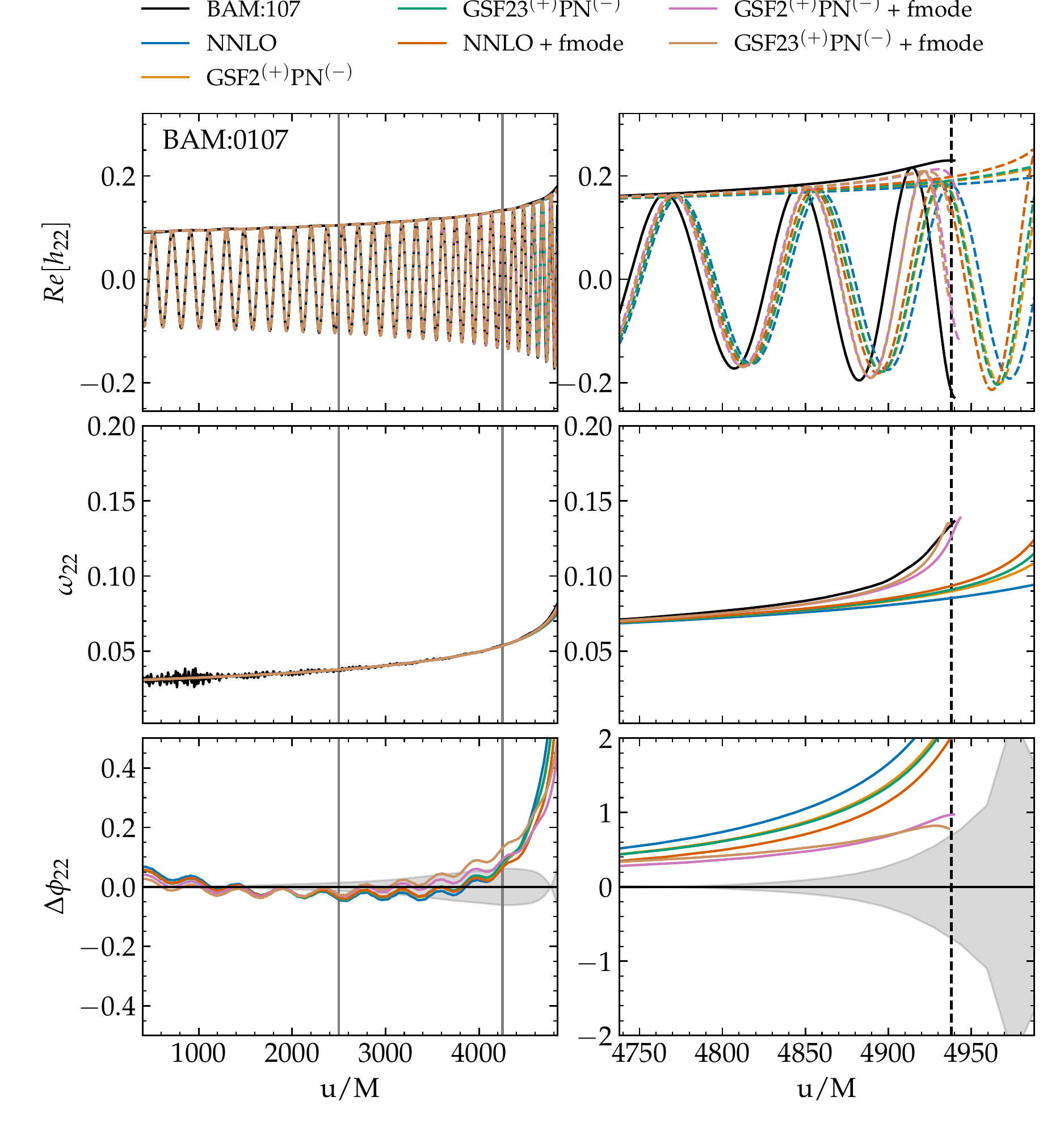}
  \caption{Waveforms (top panels), frequency evolution $\hat\omega_{22}$ (middle panels) and EOB/NR phase 
  difference $\Delta \phi^{\rm EOB/NR}$ (bottom panels) for all the non spinning BNS simulations
  considered in Fig.~\ref{fig:EbjT_qcs}. 
  The GSF+$f$-mode tidal model is the closest to NR for the \core{BAM}{0064} and \core{BAM}{107} simulations, 
  while for \core{BAM}{0037} and \core{BAM}{0095} the The NNLO+$f$-mode and the GSF models without dynamical tides 
  deliver the best waveforms.}
  \label{fig:wfs_qcs}
\end{figure*} 

The typical behaviour of the dressing factors $\alpha_{\lm}$ during the
quasi-circular merger process is shown in Fig.~\ref{fig:keff} for a
fiducial binary (that reproduces Fig.~1 of \cite{Steinhoff:2016rfi} with our
implementation). After the resonance, the
dressing factors decrease and become smaller than one or even
negative for typical BNS parameters.
Since the post-resonance behaviour is not directly modeled in the
effective Love number model it is unclear to what extent this effect is physical. 
However, given that the resonances happen before merger, this trend affects
the accuracy of the EOB waveforms that adopt this $f$-mode model.

Indeed, the behaviour of the dressing factors after the resonance can introduce unphysical features 
in the EOB dynamics by affecting the EOB light ring, $\rLR$. %, which usually increases in presence of the $f$-mode model. 
When using the PN expanded tidal model with dressed tides, the peak of the orbital frequency typically happens 
{\it after} the resonance, i.e. at $\hat{\Omega}_{\rm peak} > \hat\Omega_{f}^{(2)}$. 
Since $\hat\Omega_{\rm peak}$ (the EOB light ring) is the natural point to stop the EOB dynamics, the earlier 
resonance generates an unphysical steep increase of the waveform's amplitude approaching merger. In order to 
minimize this behaviour, the EOB model of \cite{Hinderer:2016eia, Steinhoff:2016rfi} terminates the EOB dynamics 
at the NR merger using the quasiuniversal fits of \cite{Bernuzzi:2014kca,Bernuzzi:2014owa} for which $\hat\Omega^{\rm NR}_{\rm mrg}<\hat{\Omega}_{\rm peak}$. We follow here the same procedure, but emphasize that this solution is 
not satisfactory since a well designed EOB model should not break before its light ring (this is true for 
\TEOB{} even in the binary black hole case).
Further, we manually impose that $\alpha_{\lm} \geq 1$ post-resonance.

\section{Comparison with numerical-relativity data}
\label{sec:comp}

We contrast different EOB tidal models to selected NR
simulations considering both gauge-invariant energy-angular momentum
energetics \cite{Damour:2011fu} and the $\ell=m=2$ waveform mode
phasing. We consider the NNLO \cite{Bernuzzi:2012ci}, 
\gsftides{2} \cite{Bini:2014zxa,Bernuzzi:2015rla},
and \gsftides{23} \cite{Akcay:2018yyh} prescriptions for the EOB tidal
potential with and without the
$f$-mode resonance model described above. We consider NR data from
quasi-circular and highly eccentric mergers computed respectively in
Ref.~\cite{Dietrich:2017aum,Dietrich:2017xqb} and Ref.~\cite{Chaurasia:2018zhg} using Jena's BAM code
\cite{Brugmann:2008zz,Thierfelder:2011yi}. 
The binding energy $E_b$ and the specific angular momentum $j$ are computed as described in
\cite{Damour:2011fu,Bernuzzi:2012ci}. The tidal contribution
$E_b^{\rm T}$ to the
energy curves is isolated by subtracting the relative
binary-black-hole contribution as described in
\cite{Bernuzzi:2012ci,Bernuzzi:2013rza,Bernuzzi:2015rla}. For the NR
data we use the equal mass, nonspinning binary-black-hole SXS simulation {\tt{SXS:BBH:0002}}.
For time-domain waveforms comparison, the arbitrary time and phase
relative shifts are determined by minimizing the phase
difference $\Delta\phi^{\rm EOBNR}$ over a fixed time interval $\Delta
t$, e.g.~\cite{Bernuzzi:2011aq,Dietrich:2019kaq}.

\subsection{Quasi-circular mergers}
\label{sbsec:circ}

\begin{figure*}
  \centering 
  \includegraphics[width=0.49\textwidth]{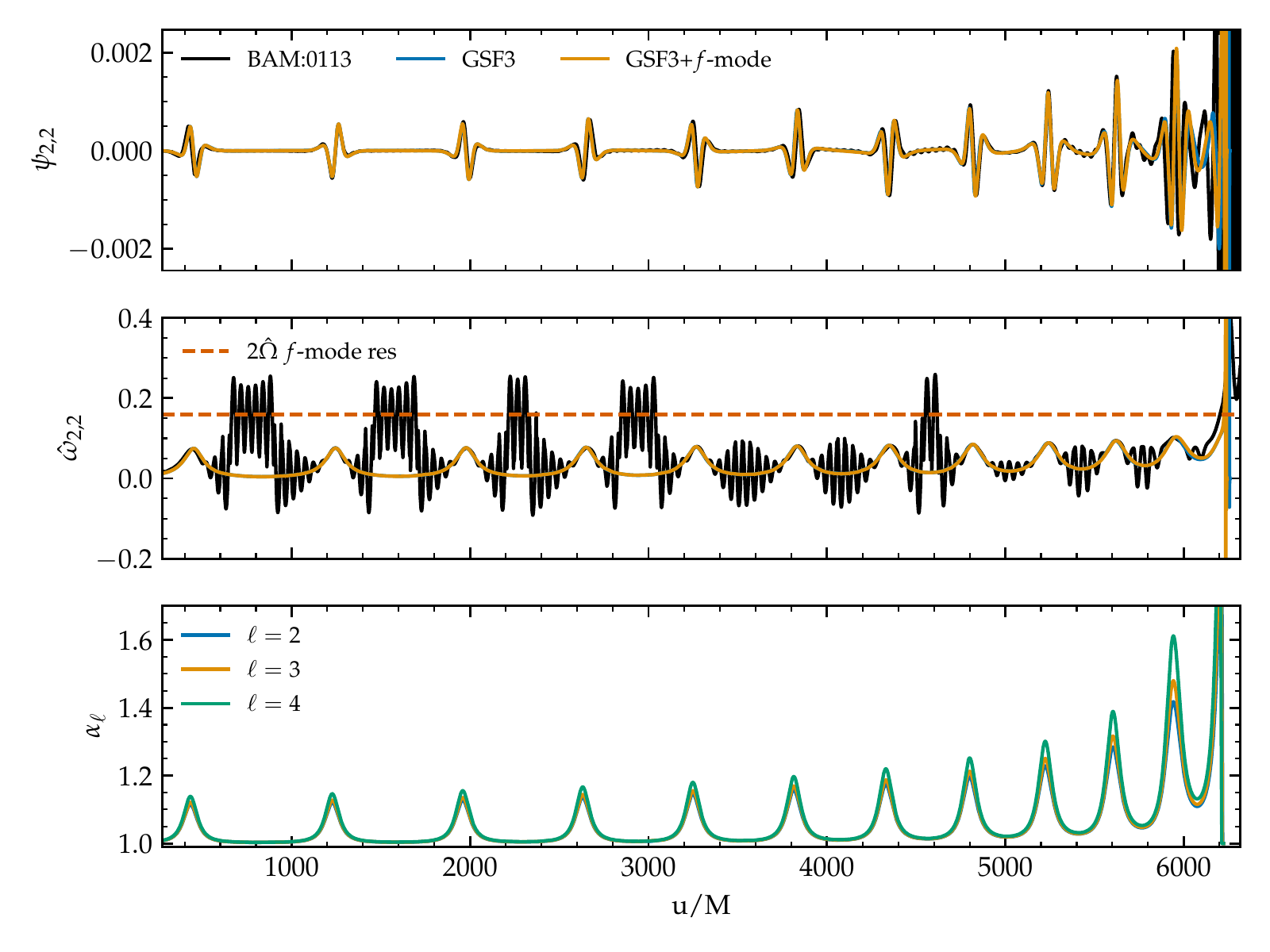}
  \includegraphics[width=0.49\textwidth]{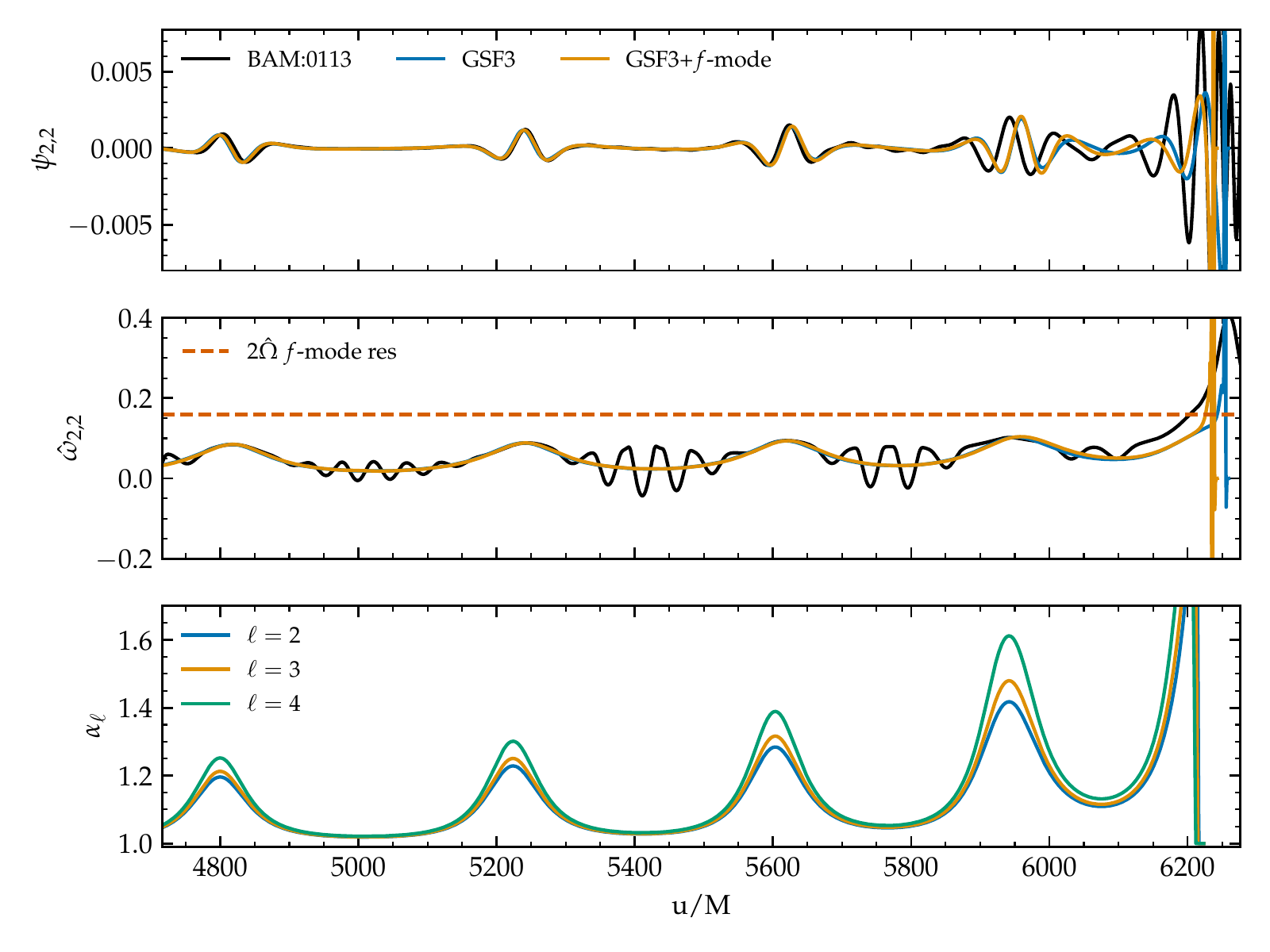}
  \caption{EOB/NR comparison between the multipolar Weyl scalars $\psi_{2,2}$ (top panel) and 	their respective frequency evolutions $\hat\omega_{22}$ (middle panel). The $f$-mode 		
  excitations  that are typically observed in NR simulations between two close encounters
  are not captured by the $f$-mode resonance model which prescribes
  $k_\ell \mapsto \alpha_\ell k_\ell$.
       }
  \label{fig:wfs_ecc}
\end{figure*}

We consider the simulations of the CoRe collaboration named
\core{BAM}{0037} \cite{Dietrich:2017aum}, \core{BAM}{0064} \cite{Dietrich:2017aum}, \core{BAM}{0095}
\cite{Dietrich:2017aum} and \core{BAM}{0107} \cite{Dietrich:2017xqb} corresponding to
nonspinning mergers with $\kappa_2^{\rm T}=187,287,73,136$ and $q = 1,1,1,1.224047$
respectively. These data is computed at multiple resolutions and show
convergent properties that allowed a clear assessment of the
errobars \cite{Akcay:2018yyh}. Hence, these are some of the most
challenging NR waveforms to reproduce with analytical models.

Figure~\ref{fig:EbjT_qcs} shows the tidal contribution to the binding
energy for the NR data and for all the considered models (top panels)
and the differences
$\Delta E_b^{\rm T} = E_b^{\rm T\,EOB} - E_b^{\rm T\,NR}$ (bottom panels). 
The EOB model based on the NNLO PN expansion of the $A_T$ potential
significantly understimates the actual tidal interaction, as it is well
known from previous results
\cite{Bernuzzi:2012ci,Bernuzzi:2015rla}. Augmenting the NNLO model
with $f$-mode resonance terms improves the agreement with NR but the
energetics are compatible only for
\core{BAM}{0095} while for the other three binaries the disagreement
remains significant. 
Further, without forcibly stopping the evolution at the NR merger (see Sec.~\ref{sec:model})
the amplitude of the waveform, too, would be largely overestimated near merger.
Note the NNLO+$f$-mode is the model employed in {\SEOB}
\cite{Hinderer:2016eia,Steinhoff:2016rfi,Steinhoff:2021dsn}.
The \gsftides{2} and \gsftides{23} models behave very similarly to
the NNLO+$f$-mode model, improving the NNLO behaviour but also
departing from the NR data for \core{BAM}{0064} and \core{BAM}{0107}.
The \gsftides{23} is currently the default choice in {\TEOB}
\cite{Bernuzzi:2015rla,Akcay:2018yyh}. If these GSF-model are
augmented with the $f$-mode the dynamics becomes too attractive and
departs from the NR data in all the considered binaries but \core{BAM}{0107} and \core{BAM}{0064}.

Figure~\ref{fig:wfs_qcs} shows the GW phasing analysis for all the simulations considered; the top, middle and bottom panel show the evolution
of the waveform's amplitude, the waveform's frequency and the phase
differences $\Delta\phi =\phi^{\rm EOB}-\phi^{\rm NR}$ respectively.
For \core{BAM}{0107}, the frequency evolution of the NNLO model significantly differs out of the
alignement interval and is not sufficiently rapid to follow
the NR data. This is in agreement with the relative energetics discussed above.
The NNLO+$f$-mode and the GSF (without $f$-mode) models improve over
the NNLO phasing but, again, the frequency evolution remains too slow
to capture the NR tides. On the contrary, the GSF2+$f$-mode models 
describe very closely the frequency evolution of the NR data,
and give the best approximation of the waveform for this binary. 
This behaviour is consistent with what
observed in the energetics above, although the merger 
-- approximated by the EOB light ring -- is reached too early
in the coalescence.

The \core{BAM}{0037} and \core{BAM}{0064} phasing analyses are qualitatively analogous 
to one another, and no model is able to reproduce the NR frequency 
evolution, although the phase of the corresponding waveform might fall within the 
NR error. Tidal effects are too attractive for $f$-mode augmented GSF models, 
and not attractive enough for the remaining models.

Differently from the others, for the \core{BAM}{0095} simulation 
the NNLO+$f$-mode and the GSF (no $f$-mode) models
are the closest to the NR data and within the error bars. 
In this case, the GW phasing analysis is compatible with the results of the energetics.

The results discussed above highlight that establishing the presence
of $f$-mode resonances in quasi-circular merger computed in numerical
relativity is not straigthforward. On the one hand, the inclusion of
this interaction in EOB models can help obtaining analytical waveforms
more faithful to NR, at least for some binaries.
This is evident in the analysis of the
equal-mass, non-spinning merger \core{BAM}{0095}, where the inclusion
of the $f$-mode resonance in the NNLO EOB model shows an excellent
agreement to NR data in both energetics and phasing as opposed to 
the NNLO EOB baseline. 
On the other hand, the $f$-mode resonance does not capture well the
waveforms of other binaries and the EOB/NR waveform agreement does not
always correspond to an improvement of the energetics (i.e. the Hamiltonian).
For example, the NNLO+$f$-mode model does not perform uniformly well with
the other equal-mass, non-spinning binaries.
The GSF+$f$-mode models, instead, give a very attractive
interaction close to merger and significantly depart from NR 
for case studies \core{BAM}{0037} and
\core{BAM}{0095}.

\subsection{Highly eccentric encounters}
\label{sbsec:hyp}

We consider the \core{BAM}{0113} simulation of Ref.~\cite{Chaurasia:2018zhg}, where 
constraint satsfying initial data are prepared and evolved for a highly
eccentric ($e_{\rm NR} \sim 0.45$) merger. The binary undergoes eleven periastron passages
before merging; each passage is characterized by a burst of GW
radiation, as shown in Fig.~\ref{fig:wfs_ecc}. Between each burst, the GW shows
oscillations compatible with the axisymmetric $f$-mode of the
(nonrotating) NS component. The oscillation frequency can be
identified also in the fluid density and it is triggered by the close 
passage to the companion \cite{Gold:2011df}. 

{\TEOB} can model these type of mergers \cite{Chiaramello:2020ehz,Nagar:2020xsk,Nagar:2021gss}. Although previous works focused on BBH systems, the extension of {\TEOB} to eccentric and hyperbolic binaries including NSs is straightforward, and we have it implemented in this work. %
The EOB/NR comparison with these type of NR data requires to fine-tune the EOB initial conditions because no analytical map is known between EOB and the initial data employed in the simulation \cite{Tichy:2012rp}. 
In order to reproduce the NR waveform, we fix the NS masses and quadrupolar tidal parameters to those employed in the NR simulation and vary independently the nominal EOB eccentricity and initial frequency until an acceptable EOB/NR phase agreement is found. This procedure is equivalent to fixing the initial frequency of the waveform and varying independently the mean anomaly and the eccentricity of the system. For this work we do not implement a minimization procedure, we instead find that manually tuning the parameters to $\hat\omega_0 = 0.0058$ and $e^0_{\rm EOB} = 0.58$ is enough to obtain a good visual EOB/NR agreement that is sufficient for our purposes.

The waveform comparison is performed in terms of the multipole $\psi_{22}=\ddot{h}_{22}$ of the Weyl pseudoscalar $\Psi_{4}$ since this quantity best highlights the $f$-mode oscillations between the bursts. As shown in the top and middle panels of Fig.~\ref{fig:wfs_ecc}, the EOB $\psi_{22}^{{\rm EOB}}$ closely matches the NR data in both amplitude and phase showing an excellent agreement during the ten periastron passages and up to merger. However, the middle panel also shows that the EOB $f$-mode model does not capture the high frequency oscillations in the GW frequency $\hat\omega_{22}$. This might not be surprising:
as shown in the bottom panel of Fig.~\ref{fig:wfs_ecc}, the $f$-mode model prescribes significant variations of the dressing factors only around the peaks of the (orbital) frequency while it is close to one in-between the peaks. By contrast, in the NR data the high-frequency oscillations are observed mainly at times between two close passages
\footnote{We note that in order to correctly compute the $f$-mode 
induced amplitude oscillations, Ref.~\cite{Chaurasia:2018zhg} corrected the multipoles for displacement-induced mode mixing \cite{Boyle:2015nqa}. Although we mainly focus on the frequency of the waveform, rather than the amplitude, Ref.~\cite{Chaurasia:2018zhg} suggests that this quantity too might be influenced by such an effect, which we do not account for here.}.
Further, by comparing the orbital frequency evolution to the resonant frequency condition, we observe that throughout the inspiral the resonance is never fully crossed.
To model this behaviour, it seems necessary to consider more complex models which consider the post-resonance dynamics -- not included in our effective model -- and for which the tidal response is evolved together with the orbital dynamics of the system \cite{Turner:1977b, Yang:2018bzx,Parisi:2017kgx} and incorporate those models in the EOB.

\begin{figure}
  \centering 
  \includegraphics[width=0.5\textwidth]{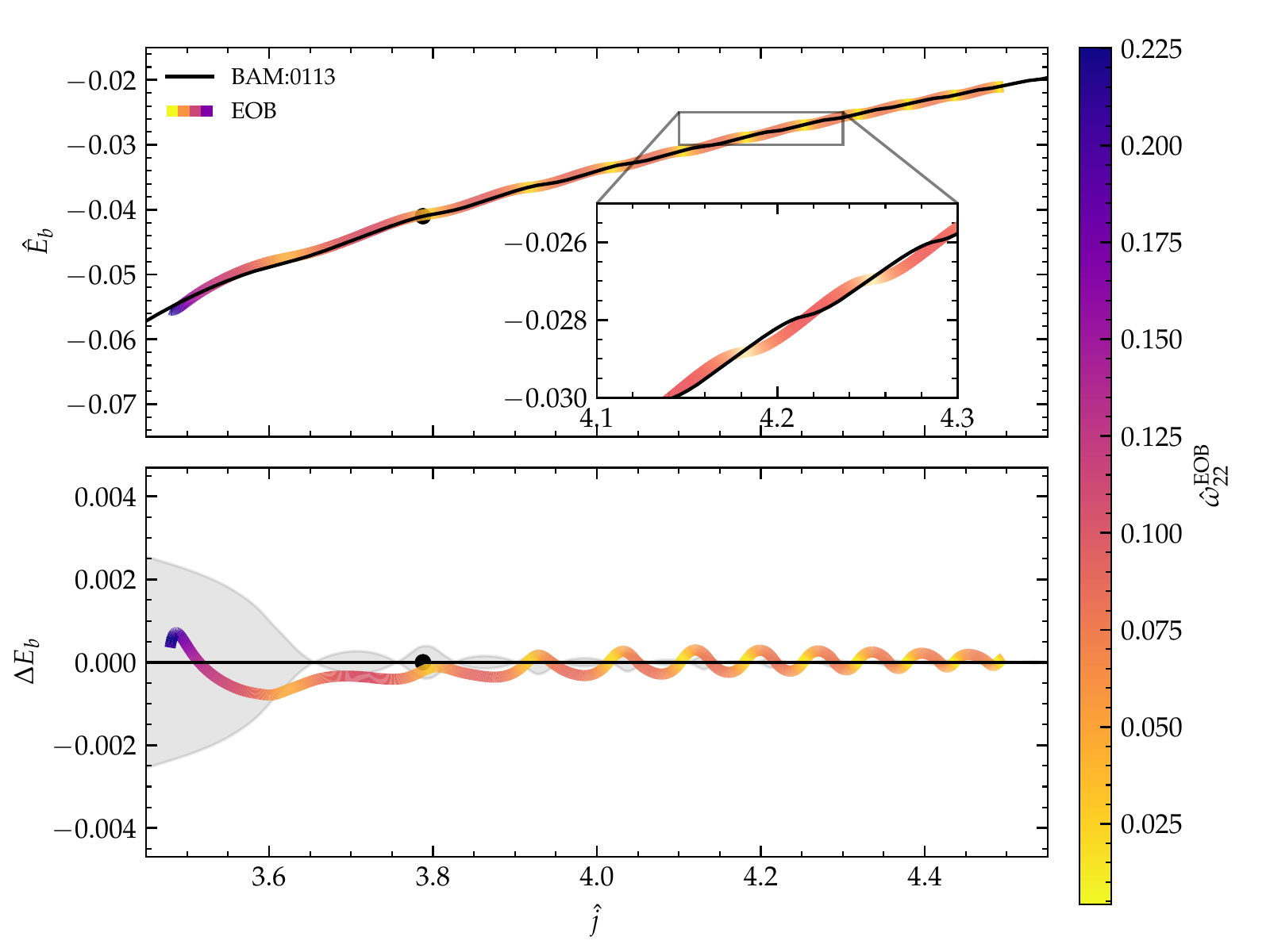}
  \caption{EOB/NR comparison between the evolution of the binding
    energy $\hat{E}_b$ of the system as a function of its orbital
    angular momentum $\hat{j}$ for the same binary of
    Fig.~\ref{fig:wfs_ecc}. The colorbar additionally indicates the EOB
    frequency evolution along the dynamics.
    The energy difference that we obtain is $\mathcal{O}(10^{-4})$, 
    compatible with the estimates of the energy carried by $f$-mode 
    oscillations via Eq. 17 of \cite{Chaurasia:2018zhg}.}
  \label{fig:dyn_ecc}
\end{figure}

We complement the GW phasing analysis with a discussion on the 
energetics. Figure~\ref{fig:dyn_ecc} shows the binding energy of the 
highly eccentric system as a function of the orbital 
angular momentum. The decrease of $\hat{E}_b(\hat{j})$ presents clear 
oscillations that can be reconducted to the close encounters. 
During each passage both $\hat{E}_b$ and $\hat{j}$ decrease
but the times at which the two NSs are apart are
characterized by approximate ``plateaus'' (moments of approximately constant energy
and angular momentum, see the inset). 
From this interpretation, it appears that the EOB and NR curves, although close, 
are not perfectly compatible: the encounters do not always align in the $\hat{E}_b(\hat{j})$
curves.
Finally, we stress that, modulo the small $f$-mode
feature, our EOB waveforms \textit{quantitatively} reproduce highly
eccentric NR simulations up to few orbits before merger. Ours is the
first EOB model capable of describing highly eccentric comparable-mass
system including neutron stars, and this is to our knowledge the first
EOBNR comparison of this kind. The striking agreement between EOB and NR in Fig.~\ref{fig:wfs_ecc} attests to the goodness of the radiation reaction model employed within \TEOB. 

\section{Model selection on GW170817} 
\label{sec:GW170817}

We now apply our models to GW170817, using  the \bajes{} 
pipeline \cite{Breschi:2021wzr} and the {\tt dynesty} \cite{Speagle:2020} sampler.
We consider $128$ seconds of data around GW170817 GPS 
time, and analyze frequencies between $23$ Hz and $2048$ Hz.
The employed prior is uniform in component masses and tidal parameters, isotropic in spin components 
and volumetric in luminosity distance. It spans the ranges of chirp mass $\mathcal{M}_c \in [1.1, 1.3]$, mass ratio
$q \in [1,3]$, spin magnitudes $\chi_i \in [0, 0.05]$, tidal parameters $\Lambda_i \in [0, 5000]$
and distance $D_L \in [20, 100]$ Mpc.
We consider three models for our analysis: the \gsftides{23} model, the 
the \gsftides{23} model augmented with dynamical tides and the NNLO model, also augmented with dynamical tides\footnote{For computational convenience we do not employ dressed spin-quadrupole 
parameters}. 
When using the $f$-mode resonance model, 
we either fix the values of $\bar{\omega}^{(2)}_{f A}$, $\bar{\omega}^{(2)}_{f B}$ via the 
quasi-universal relation of
\cite{Chan:2014kua} or we infer them independently of $\Lambda$, imposing 
uniform priors on $\bar{\omega}^{(2)}_{f i} \in [0.04, 0.14]$ with $i = A,B$.

\begin{figure}
  \centering 
  \includegraphics[width=0.5\textwidth]{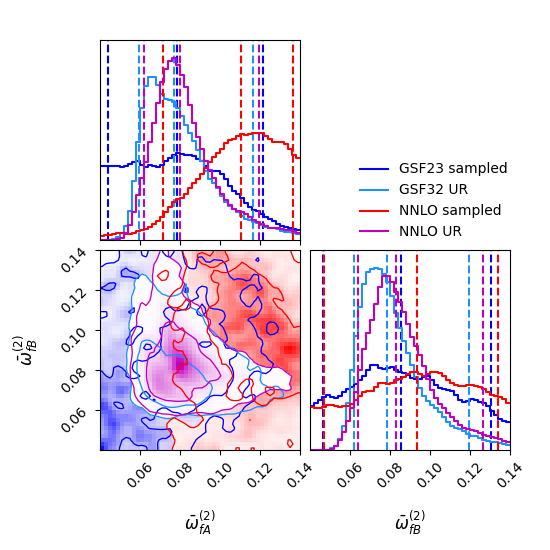}
 \caption{Posterior samples for $\bar{\omega}^{(2)}_{f A}$ and $\bar{\omega}^{(2)}_{f B}$ extracted from 
 GW170817 via direct sampling (blue, red) or by applying quasi-universal relations to the mass and tidal parameters recovered (cyan, magenta). The sampled values span the entire interval investigated, indicating that we are not able to precisely extract the $\ell=2$ resonance frequency from GW170817 data.}
  \label{fig:pe_gw1710817_bomgf}
\end{figure}

Posteriors for the intrinsic parameters can be inspected in App.~\ref{app:posplot}.
The evidences recovered with the five models are instead reported in Tab.~\ref{tab:evidence}.
The data mildly favors the GSF tidal model and the GSF model augmented by $f$-mode resonances
with respect to the NNLO dynamical tides model.
When sampling the resonance frequencies (see Fig~\ref{fig:pe_gw1710817_bomgf}), we find that 
it is not possible to precisely determine $\bar{\omega}^{(2)}_{f}$ from GW170817 data. For both tidal baselines, 
the recovered $\bar\omega^{(2)}_B$ distribution is consistent with the flat prior imposed. The distribution of
$\bar\omega^{(2)}_A$, instead, allows only to impose an upper or lower bound on the 
$f$-mode frequency, depending on whether the GSF or NNLO tidal model is employed.
This is consistent with what previously observed in Ref.~\cite{Pratten:2021pro}:
it is not possible to accurately determine $\bar{\omega}^{(2)}_{f}$ from GW170817 data. 

 \begin{table}[t]
   \centering    
   \caption{Logarithmic evidences $\log Z$ of the five models employed in our GW170817
    reanalysis and their Bayes' factors computed against the GSF model. 
    The GSF model is slightly favoured over the GSF + $f$-mode model 
    both when we do and do not attempt to infer $\bar{\omega}_{f}^{(2)}$ from the GW data. 
    NNLO + $f$-mode models, instead, appear mildly  
    disfavored with respect to those employing the adiabatic GSF tidal baseline.}
   \begin{tabular}{ccc}        
     \hline
     Model (X) & $\log{Z}$ & $\log{\mathcal{B}^{\rm X}_{\rm GSF}}$ \\
     \hline
     GSF                        & $480.23 \pm  0.18$ &  $0$     \\
     GSF + $f$-mode             & $479.61 \pm  0.18$ &  $-0.62$ \\
     GSF + $f$-mode + sampling  & $479.57 \pm  0.18$ &  $-0.66$ \\
     PN  + $f$-mode + sampling  & $479.22 \pm  0.18$ &  $-1.01$ \\
     PN  + $f$-mode             & $479.15 \pm  0.18$ &  $-1.08$ \\

     \hline
   \end{tabular}
  \label{tab:evidence}
\end{table}

\begin{figure}
  \includegraphics[width=0.44\textwidth]{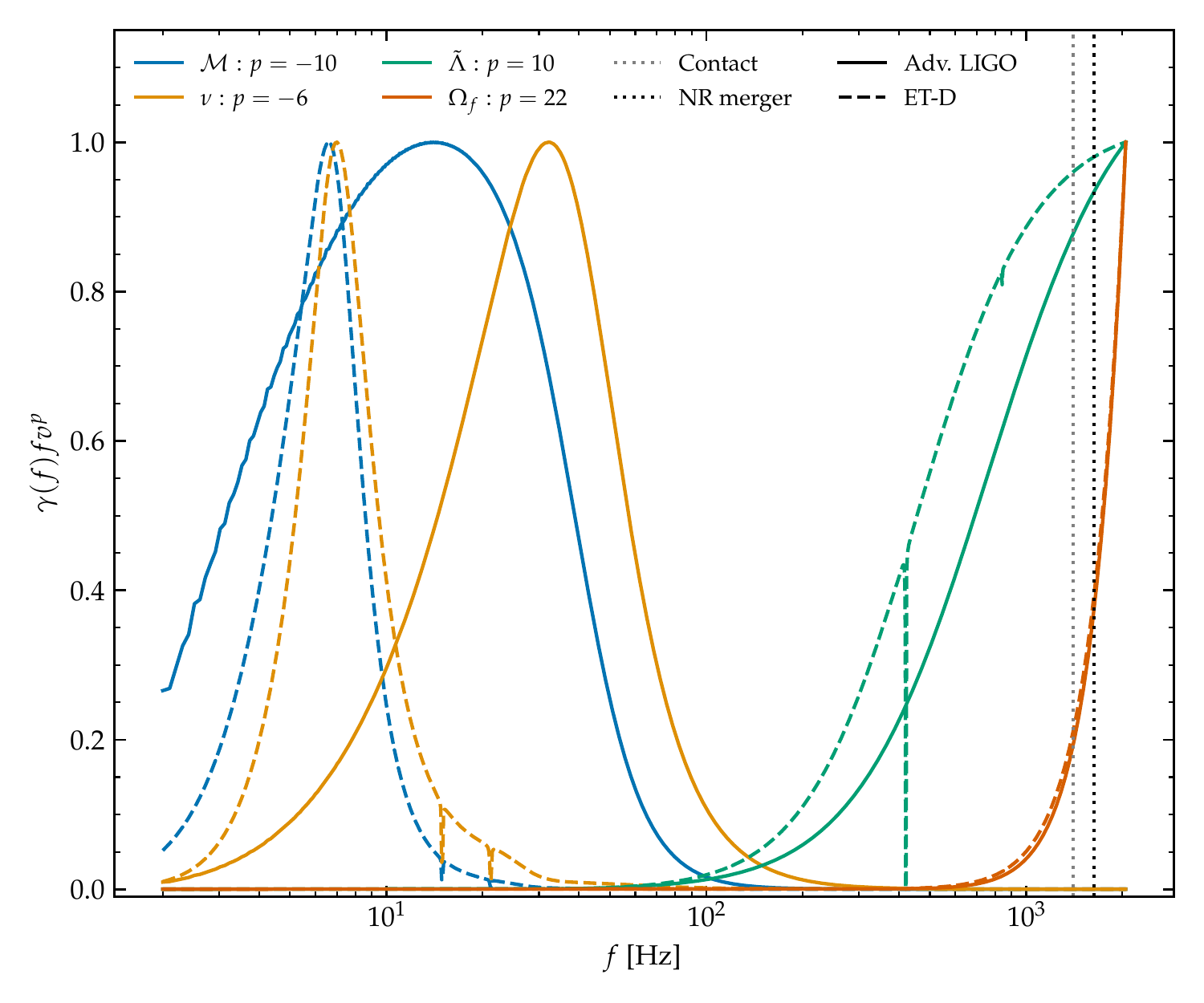}
  \caption{Fisher matrix integrands, computed as in \cite{Damour:2012yf}, evaluated considering the leading order phase contribution for chirp mass, symmetric mass ratio, effective tidal parameter and $f$-mode frequency. Straight curves are computed using advanced LIGO PSD \cite{aLIGODesign_PSD}, while dashed lines are estimated with Einstein Telescope PSD \cite{Hild:2010id}.
Notably, the $f$-mode resonance frequency is informed mainly by very high frequencies, larger than merger or contact.}
  \label{fig:fm_int}
\end{figure}

A simple Fisher Matrix study (Fig.~\ref{fig:fm_int}) immediately 
clarifies the reason for the fact illustrated above. 
Following Ref.~\cite{Damour:2012yf}, we compute the diagonal Fisher Matrix (normalized)
integrands $\gamma(f) f v^{p}$:
\begin{equation}
\gamma(f) = \frac{f^{-7/3}/S_n(f)}{\int_{\rm fmin}^{\rm fmax} f^{-7/3}/S_n(f)} \, ,
\end{equation}
where $v = (M \pi f)^{1/3}$ and $p$ depends on the parameter considered.
Employing the PN frequency domain model of Ref.~\cite{Schmidt:2019wrl} and considering 
only the leading order for each of the studied parameters, one finds that $p_{\mathcal{M}} = -10$, 
$p_{\nu} = -6$, $p_{\tilde\Lambda} = 10$ and $p_{\bar{\omega}_f} = 22$.
This indicates that $f$-mode parameters are determined close to the resonance frequency, 
where the effect of the model is strongest ($\alpha_\ell > 1$).
For GW170817, such frequencies are dominated by the detectors' noise.

\begin{figure}
  \centering 
   \includegraphics[width=0.45\textwidth]{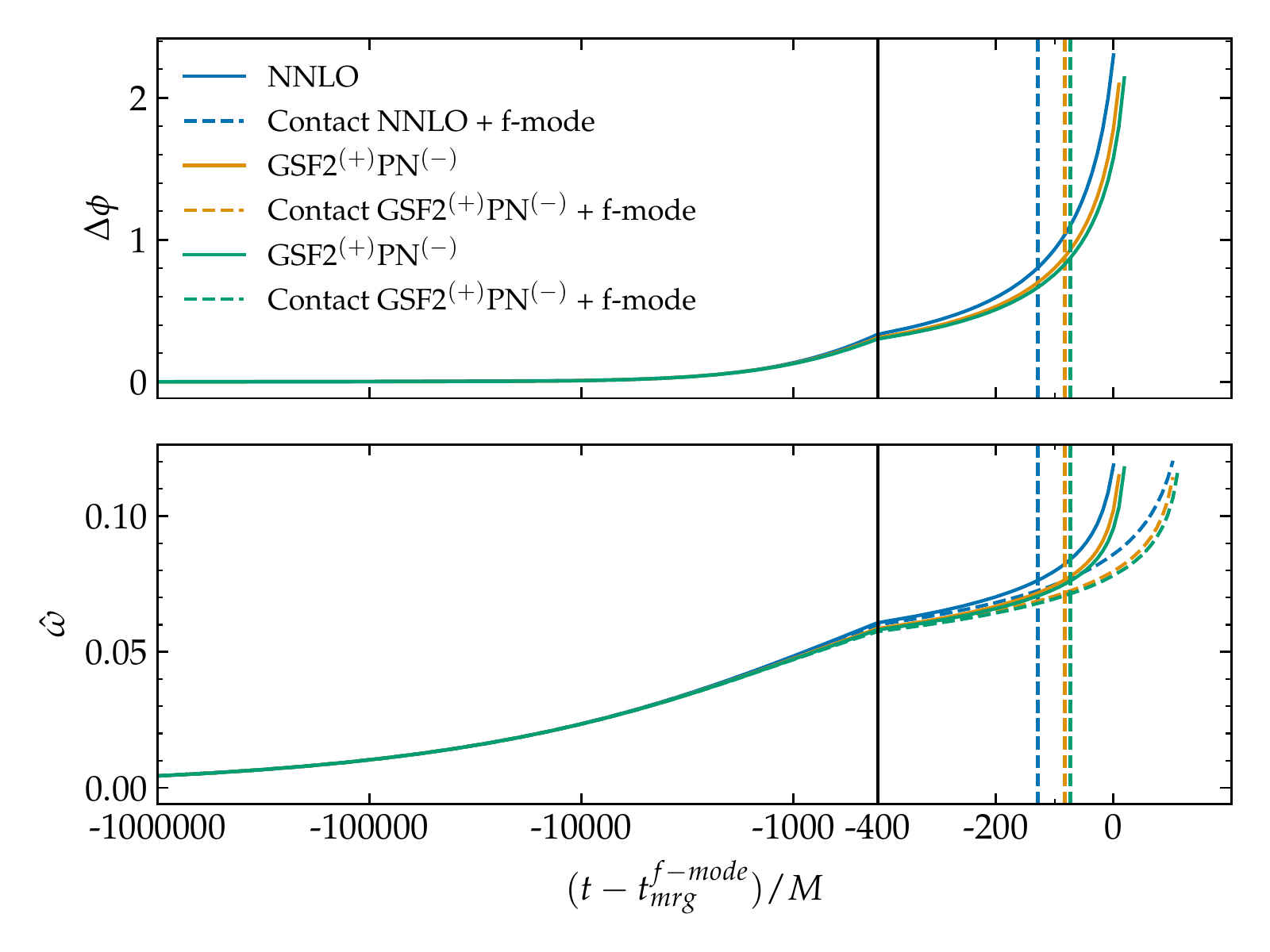}
 \caption{Phase difference $\Delta\phi = \phi^{\rm f-mode} - \phi^{\rm no~f-mode}$ due to dynamical tides
 for a target binary system with $M=2.7 M_{\odot}$, $\Lambda_1 = \Lambda_2 = 978$ and $q=1$
 from a starting frequency of $20$ Hz. We employ three different baseline tidal models: 
 NNLO (blue), \gsftides{2} (orange) and \gsftides{23} (green). 
 Dashed colored lines indicate time of contact between the two stars. Note that the horizontal axis scale changes at $(t - t_{\rm mrg})/M = -400$.}
  \label{fig:acc_phase}
\end{figure}

It is worth noting that in the region where dynamical effects become more prominent 
(above the frequency of contact between the two stars),
the model itself is not physically grounded.
PE studies such as the ones presented in 
\cite{Pratten:2019sed, Pratten:2021pro, Williams:2022vct} circle this issue by 
generating waveforms exclusively up to the contact frequency, measuring
the secular accumulated phase difference due to the effect of dynamical tides away from the resonance over
a very large number of GW cycles. 
Directly testing the physical validity of this approach is challenging, as it would require 
extremely long NR simulations which are currently unavailable.
Within our EOB model, we find an accumulated phase difference due to dynamical tides of $\sim 0.5-1$~rad at contact
for a $1.35+1.35$ reference binary from $20$ Hz (Fig.~\ref{fig:acc_phase}).
Most of the phase is accumulated above $300$ Hz.
This phase difference might become measurable with third generation detectors \cite{Williams:2019vub}, 
although biases due to an imperfect knowledge of the point mass
and (adiabatic) tidal sectors of the models
could affect future measurements.

\section{Conclusions}
\label{sec:conc}

In this paper we studied the $f$-mode resonances in three different
tidal flavors of the EOB approximant \TEOB, namely the NNLO PN, the
GSF2 and the GSF23 resummed models. We performed detailed EOB/NR comparisons
of both waveforms and gauge-invariant energetics focusing on four selected high-resolution
simulations of quasi-circular and one highly eccentric BNS merger.

In the circular merger case, we found that the NNLO+$f$-mode model performs similarly to the
GSF-resummed ones without $f$-mode, and that -- in all but one case --
the model fails to capture either the energetics or the waveform of NR
data. 
Therefore, while the studied model certainly represents a viable alternative to GSF
resummation, we suggest caution when trying to extract physical information from it 
via GW data analysis of real events: no clear signature of the presence of $f$-mode 
resonances after the NS contact can be assessed from NR simulations.

In the highly eccentric case, we found that the effective $f$-mode
model does not capture the oscillations in the NR data
(small oscillations in amplitude and frequency of
the waveform centered around the proper mode star frequency).
This is not unexpected because (i) the $f$-mode model considered here was
specifically derived for quasi-circular orbits, (ii) the resonant
condition is not met during the close passages.
Aside from the small oscillation feature, we demonstrated that our 
\TEOB~ for generic orbits quantitatively reproduces the NR waveform and frequency
evolution with high-accuracy up to merger. To our knowledge, this is
the first EOB/NR comparison of highly eccentric BNS mergers.

Finally, we applied our adiabatic- and dynamic- tidal models to 
GW170817, and found that models which employ the GSF tides baseline
are mildly preferred over models that employ NNLO dynamical tides.
Additionally, we were not able to determine $f$-mode resonances from the GW
data. The reason for this was immediately understood 
in terms of a simple Fisher Matrix study, which highlights 
that dynamical tides are effectively measured at very high frequencies ($\geq 1$ kHz). 

Our results should be considered when adopting $f$-mode resonance
models in gravitational-wave analyses and parameter estimation. Some
of these analyses might be carried with phenomenological models that
can reproduce some waveform features but do not allow for a careful check of the
underlying dynamics and Hamiltonian \cite{Schmidt:2019wrl,Andersson:2019dwg}. 
A careful validation of these models against more complete EOB models
appears necessary in the future.

\begin{acknowledgments}
  We thank Jan Steinhoff, Huan Yang, William East, Aaron Zimmerman and Nathan Johnson-McDaniel 
  for discussions during the preparation of this manuscript, and Jacopo Tissino for pointing us to 
  the correct ET-D PSD.
  RG is supported by the Deutsche Forschungsgemeinschaft (DFG) under Grant No.
  406116891 within the Research Training Group RTG 2522/1.
  SB acknowledges support by the EU H2020 under ERC Starting
  Grant, no.~BinGraSp-714626.  
  The authors will always be indebited to Beppe Starnazza.
  SB acknowledges the hospitality of KITP at UCSB and partial support by the National Science Foundation under Grant No. NSF PHY-1748958 during the conclusion of this work.
  \TEOB{} ``GIOTTO'' is publicly available at
  
  {\footnotesize \url{https://bitbucket.org/eob_ihes/teobresums/src/master/}}
  
\end{acknowledgments}

\onecolumngrid
\appendix

\section{Effective Love number model for $f$-mode resonances} 
\label{app:keff}

This appendix summarizes the effective Love number model introduced in
\cite{Hinderer:2016eia,Steinhoff:2016rfi}.
The model results from an approximate
solution of the equation definining the effective quadrupolar Love number
\be\label{eq:keff_def}
k^{\rm eff}_2 = \frac{E_{ij}Q^{ij}}{E^2}\,, 
\ee
for a Newtonian inspiral. In Eq.~\eqref{eq:keff_def},
$E_{ij}=\p_i\p_j\phi$ is the external quadrupolar field derived from
the Newtonian potential $\phi$ and $Q_{ij}$ the NS's quadrupole.
The Love number $k_\ell$ of star $A$ (or analogously 
the tidal polarizability parameter) is substituted with an effective
Love number that depends on the orbital frequency and its $f$-mode
frequency $\bar{\omega}_f^{(\ell)}$,  
\be\label{eq:keff}
k_\ell\mapsto k_\ell^{\rm eff}:=\alpha_{\ell m}(\nu,\Omega,\bar{\omega}_f^{(\ell)},X_{\rm A})k_\ell \,.
\ee
The enhancement, or dressing, factor $\alpha_{\ell m}$ in Eq.~\eqref{eq:keff} is a
multipolar correction valid for $\ell=m$ and given by
\begin{align}\label{eq:alpha}
  \alpha_{\ell m} &=
a_\ell+b_\ell\,\left\{
\frac{x^2}{x^2-1} + \frac{5}{6}\frac{x^2}{1-x^{5/3}}
+\frac{x^2}{\sqrt{\epsilon}} \left[
  \cos\left(\Omega'\hat{t}^2\right)
  \int_{-\infty}^{\hat{t}}\sin\left(\Omega' s^2\right)\d s
  - \sin\left(\Omega'\hat{t}^2\right)
  \int_{-\infty}^{\hat{t}}\cos\left(\Omega' s^2\right)\d s
  \right]
  \right\}\,.
\end{align}
In the above equations, the first multipolar coefficients are
$(a_2,a_3,a_4)=(1/4,3/8,29/64)$,
$(b_2,b_3,b_4)=(3/4,5/8,35/64)$; the multipolar parameter
\be
x := x_m = \frac{\bar{\omega}_f^{(\ell)}}{m\hat\Omega X_A}\,
\ee
controls the frequency of the $f$-mode resonance, $\Omega'=3/8$ and 
\be
\epsilon := \frac{256}{5}\nu\left(\frac{\bar{\omega}_f^{(\ell)}}{mX_A}\right)^{5/3}
\ , \ \ \
\hat{t} := \frac{8}{5}\frac{1}{\sqrt{\epsilon}}\left( 1 - x^{5/3} \right)\,.
\ee
The first two terms in Eq.~\eqref{eq:alpha} are singular at the
resonance ($x=1$). The integrals in the third term reduce to
Fresnel integrals
\be
F_S(z) := \sqrt{\frac{2}{\pi}} \int_0^z\sin\left(s^2\right)\d s \,, \ \ 
F_C(z) := \sqrt{\frac{2}{\pi}} \int_0^z\cos\left(s^2\right)\d s \,,
\ee
by writing
\be
\int_{-\infty}^{\hat{t}}\cos\left(\Omega' s^2\right)\d s
% with Mathematica convention
% = \sqrt{\frac{\pi}{2\Omega'}}\left[ F_C\left(\infty\right) + F_C\left(\sqrt{\frac{2\Omega'}{\pi}}\hat{t}\right) \right]
% = \sqrt{\frac{\pi}{2\Omega'}}\left[\frac{1}{2}+F_C\left(\sqrt{\frac{2\Omega'}{\pi}}\hat{t}\right)\right]
% with other:
 = \sqrt{\frac{\pi}{2\Omega'}}\left[ F_C\left(\infty\right) + F_C\left(\sqrt{\Omega'}\hat{t}\right) \right]
 = \sqrt{\frac{\pi}{2\Omega'}}\left[\frac{1}{2}+F_C\left(\sqrt{\Omega'}\hat{t}\right)\right]
 \ee
and similarly for the other.

In \TEOB{}, the dressing factors are computed along the dynamics and
using the circular frequency $\hat\Omega = u^{3/2}$.
Following {\tt LAL}'s \SEOB{} implementation of Steinhoff~\footnote{\url{https://github.com/jsteinhoff/lalsuite/tree/tidal_resonance_NSspin}}, the singular
terms in Eq.~\eqref{eq:alpha} are substituted by their expansion near
$x=1$ if $x-1<10^{-2}$. The tidal coupling constants are then calculated with the
dressing factors and used with any prescription for the EOB tidal
potentials. This way, the $f$-mode resonant effect propagates into the
EOB dynamics. Spin interactions in \TEOB{} are modeled using the centrifugal radius~\cite{Damour:2014sva}, 
\be
r_c = \left(r^2 + a_{02}(1+2u) + \Delta_{a2}u + \left(\Delta_{a2}^{\rm NNLO}+\Delta_{a4}^{\rm LO}\right)u\right)^2\,,
\ee
which include quadrupole $S^2$ effects at NNLO and also $S^4$ terms.
The functions $a_{02}$ and $\Delta$'s contain the quadrupole moments $C_{Q_A}$ of the NS; $\Delta_{a4}^{\rm LO}$ also contains the ocutupole and hexapole parameters.
The $C_{Q_A}$ (and their derivatives) are computed using the fits of
Ref.~\cite{Yagi:2013awa} using the dressed tidal
polarizability parameters; no dressing is instead applied to the other
parameters for simplicity.  
Finally, a $\ell=m=2$ waveform's amplitude correction due to the
$f$-mode is applied using the dressing factor
\be\label{eq:hathalpha22}
\hat\alpha_{22} = x^2\frac{\alpha_{22}(1+6\frac{X_B}{x^2})-1}{3(1+2X_B)}\,.
\ee
This result was first reported in Ref.~\cite{Dietrich:2017feu} with a
different coefficient and without any detail on the calculation.
It was later reported with the coefficient used above in
Ref.~\cite{Steinhoff:2021dsn}, and it is used in this form also in
Steinhoff's {\tt LAL} implementation.

\begin{figure}
  \centering 
  \includegraphics[width=0.8\textwidth]{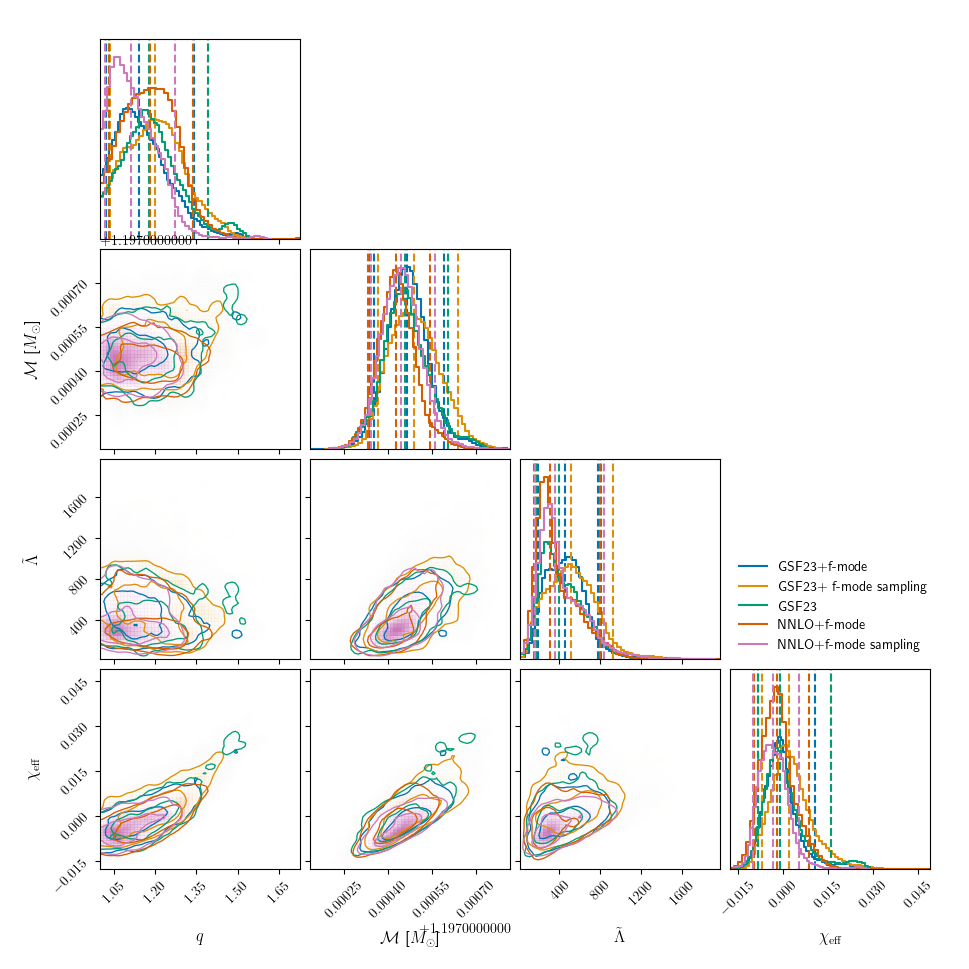}
 \caption{Marginalized, two-dimensional posterior samples for GW170817 
 obtained with the tidal flavors of {\tt TEOBResumS} listed in Tab.~\ref{tab:evidence}.}
  \label{fig:pe_gw1710817}
\end{figure}

\section{Posterior plots for GW170817}
\label{app:posplot}

in this appendix we display the posterior distributions of the intrinsic parameters 
that we obtained from our GW170817 reanalyses.
Figure \ref{fig:pe_gw1710817} displays the marginalized two-dimensional posterior samples 
for the chirp mass $\mathcal{M}$, the mass ratio $q$, the effective spin $\chi_{\rm eff}$
and the tidal parameter $\tilde{\Lambda}$.

\end{document}